\documentclass[acmsmall,natbib=false,nonacm]{acmart}
\usepackage{array}
\usepackage{multirow}
\usepackage{makecell}
\usepackage{graphicx}
\graphicspath{{figure/}}
\usepackage{url}
\usepackage{pifont}
\newcommand{\cmark}{\ding{51}}
\setcopyright{none}
\usepackage{adjustbox}

\AtBeginDocument{%
  }

\RequirePackage[
  datamodel=acmdatamodel,
  style=acmnumeric,
  ]{biblatex}

\begin{document}

\title[Demonstrably Informed Consent]{Demonstrably Informed Consent in Privacy Policy Flows: Evidence from a Randomized Experiment}


\author{Qian Ma}
\orcid{}
\affiliation{%
  \institution{The Pennsylvania State University}
  \city{}
  \state{}
  \country{USA}}
\email{qfm5033@psu.edu}

\author{Aditya Majumdar}
\affiliation{%
  \institution{The Pennsylvania State University}
  \city{}
  \country{USA}}
\email{adity@psu.edu}

\author{Sarah Rajtmajer}
\affiliation{%
  \institution{The Pennsylvania State University}
  \city{}
  \country{USA}}
\email{smr48@psu.edu}

\author{Brett Frischmann}
\affiliation{%
  \institution{Villanova University}
  \city{}
  \country{USA}}
\email{brett.frischmann@law.villanova.edu}

\renewcommand{\shortauthors}{Qian M et al.}

\begin{abstract}
Privacy policies are meant to inform people about how their data will be collected, used, and shared, yet consent is often reduced to clicking ``I agree.'' We ask whether pedagogical friction--brief interventions that slow users down, focus attention, or explain important terms--can improve understanding of consequential terms without excessive burden. In a randomized experiment with 293 parents evaluating a children’s learning app, we test six interface designs, from a standard text policy to versions using highlighting, plain-language explanations, pacing, sectioning, and timed slide recap. Participants completed a comprehension quiz; in three conditions, those scoring below 80\% reviewed the policy and retook the quiz. The slide recap condition had the highest first-attempt passing rate (41.7\%), and 66.4\% of retakers improved. In ungated conditions, 97.3\% of participants below the comprehension threshold still consented. These findings suggest that targeted friction can support comprehension while highlighting the gap between recorded agreement and demonstrated understanding.
\end{abstract}





\authorsaddresses{}
\maketitle

\section{Introduction}
%Consent in privacy policy flows is not only a legal or disclosure problem but also an interaction design problem. 
Digital consent interfaces often make agreement easy while doing little to establish whether users understand what they are agreeing to. A click on ``I agree'' tells us little about whether a person actually read or understood consequential terms~\cite{nissenbaum2011contextual,daniel2013privacy}. This problem is particularly acute for privacy policies, which are long, technical, and routinely ignored~\cite{mcdonald2008cost,reidenberg2015disagreeable}. %Interfaces often position policy review 
Policy review is typically positioned as an obstacle between users and access to a service they would like to use~\cite{nouwens2020dark,DBLP:journals/popets/MachuletzB20}. As a result, consent can become a procedural action instead of an informed choice. %From an interaction design perspective, the challenge is to design a consent process that supports understanding before agreement. In this paper, we study how the design of a privacy policy consent flow can support demonstrated comprehension of a small set of substantively important terms before a consent decision.

Recent work at the intersection of privacy law and technology argues for a stronger standard of \emph{demonstrably informed consent}~\cite{frischmann2026defendingconsent}. In this view, the party seeking consent should be able to show that the person giving consent both intended to consent and understood the substantively important terms to which they agreed. Critically, this requires evidence of comprehension as well as agreement. The standard of demonstrably informed consent raises a concrete interaction design challenge: a consent process must both support users in understanding consequential terms and provide some basis for demonstrating that understanding, while keeping the additional time and effort manageable. We treat demonstrably informed consent as a design problem and ask how interface mechanisms can support and assess comprehension at the point when consent is requested.

One way to address this challenge is through carefully calibrated friction. Although friction is often treated as a usability cost, prior work shows that small, well-placed interventions can redirect attention and interrupt automatic behavior~\cite{ma2026learning,frischmann2023friction,sunshine2009crying,akhawe2013alice}. Moderate friction may also support learning and reflection, while overly frictionless interactions can work against these goals~\cite{Zohar2026}. Timing matters as well. Guidance is most useful when it appears at a moment when a person can act on it~\cite{DBLP:journals/ubiquity/Fogg02,DBLP:conf/persuasive/Fogg09a}, and timely cues can increase attention and elaboration~\cite{michie2011behaviour,petty2012communication}. Microlearning similarly uses short, focused instruction embedded within a task to support comprehension without requiring a separate training activity~\cite{hug2006microlearning,allela2021introduction,sankaranarayanan2023microlearning}. Drawing on these ideas, we use \textit{pedagogical friction} to describe small, structured interventions embedded within a consent process that deliberately introduce time, attention, or effort in order to support understanding.

We study pedagogical friction in the context of parental consent for a children’s learning app. %a privacy policy for an educational app for young children. 
In a randomized experiment with 293 parents of children ages 3-8, participants reviewed the same privacy policy using one of six interface designs that varied how key terms were presented and paced. The designs included a standard text policy as well as versions that highlighted important terms, added plain-language explanations, divided the policy into paced sections, or provided a timed slide recap. %and randomly assigned them to one of six conditions that varied presentation format and pacing: plain text privacy policy (G0); privacy policy with substantive terms highlighted (G1); highlighted policy with explanatory blurbs (G2);  privacy policy with visual presentation of key privacy terms through a timed slideshow (G3); a paced and sectioned version of G1 (G4); and a paced and sectioned version of G2 (G5). 
After reviewing the policy, all participants completed a six-question comprehension quiz. In three conditions, participants who did not meet an 80\% comprehension threshold reviewed the policy again and retook the quiz. We recorded comprehension, time spent reviewing the policy and completing the quiz, consent decisions, and participants’ perceptions of the experience. %policy review time, quiz time, quiz responses, consent decisions, and exit survey responses.
Our work addresses two research questions:

\noindent \textbf{RQ1}: How does demonstrated comprehension of important terms vary across %privacy policy review designs with 
consent interfaces that incorporate different forms of pedagogical friction?

\noindent \textbf{RQ2}: What costs accompany frictional interventions, and how do users experience those costs?%are they tolerated by users?

Our results suggest that the benefits of pedagogical friction depend on both the design and degree of intervention. The timed slide recap condition had the highest observed first-attempt threshold attainment on the comprehension quiz, followed by the paced, sectioned highlighting condition, although neither significantly outperformed the plain-text control after correction. Among participants who did not initially meet the 80\% comprehension threshold and were given the quiz a second time, 66.4\% improved their score; the paced, sectioned condition with plain-language explanations had the largest average score gain and the highest second-attempt threshold attainment. The more structured and paced interventions imposed additional policy-review time costs, with participants reporting varying levels of burden across conditions. Strikingly, in ungated conditions, 97.3\% of participants who scored below the 80\% comprehension threshold nevertheless chose to consent. Together, these findings highlight a persistent gap between recorded agreement and demonstrated understanding, while showing that different forms of friction may support comprehension at different points in the consent process.

This work makes three primary contributions. First, we introduce pedagogical friction as a design framing for consent interactions in which purposeful costs in time, attention, or effort are used to support comprehension rather than simply to slow users down. %process in consequential consent interfaces.} 
%We frame consent as an interaction that can support comprehension through structured exposure, pacing, explanation, comprehension checks, and opportunities for re-exposure, while making the resulting user burden explicit.
Second, %Empirical evidence about how pedagogical friction designs support different stages of demonstrated comprehension.}
through a randomized six-condition study with 293 participants, we provide empirical evidence about how different consent interface designs relate to initial comprehension and to recovery among participants who did not initially meet the comprehension threshold. Third, we make the tradeoff between comprehension support and interaction cost measurable by connecting comprehension outcomes with time-on-task and perceived burden, and we document a persistent gap between recorded agreement and demonstrated comprehension. %show why comprehension support must be calibrated against interaction cost. We connect comprehension outcomes with time-on-task, perceived burden, and consent behavior.

%These contributions position pedagogical friction as a reusable design framing for consent interfaces that seek to support demonstrated comprehension while accounting for interaction cost. Analysis scripts to support replication and adaptation are available at: \url{https://anonymous.4open.science/r/consent_comp/}.

\section{Related Work}

\subsection{Friction in Consent Flows}

A common pattern in contractual and privacy policy consent flows is fast progression with limited attention to terms~\cite{obar2020biggest}. More broadly, related work on prosocial friction treats small amounts of resistance as a tool to slow users down at consequential moments and to support more deliberate choices~\cite{natali2023per,frischmann2023friction,cabitza2019programmed,gruning2024framework}. Closely related ideas appear under different terms, including frictional design, friction-in-design, microboundaries, programmed inefficiencies, seamful design, slow design, and work on consent interface patterns. Across these lines of work, the shared claim is that interfaces can shape what people notice and what they do ~\cite{nouwens2020dark}.

Seamful design highlights how making system limitations visible can prompt reflection about what the infrastructure is doing~\cite{chalmers2003seamful}. Slow design argues for interfaces that create time and space for deliberation in settings where speed works against careful judgment~\cite{grosse2013slow}. Cox et al.~\cite{cox2016design} describe microboundaries as small interaction barriers that interrupt habitual use and encourage more mindful decision making. Frischmann and Benesch~\cite{frischmann2023friction} argue that well-chosen friction can support reflection and self-determination, and propose a framework that distinguishes types of friction, direct effects on subjects, and intended purposes. Evidence from security and privacy research also suggests that attention matters for outcomes. Warnings that interrupt the task and require an explicit user action tend to have stronger behavioral effects than passive cues that users can ignore~\cite{kaiser2021adapting,utz2023comparing}. 

Our study builds on this body of work by applying a pedagogical friction framing to privacy policy consent flows, shaping attention at consequential moments, teaching users about substantive privacy terms, and measuring comprehension.

 \subsection{Microlearning}
 Microlearning is a broad term for instructional approaches that deliver targeted, action-oriented content in small units designed to achieve specific objectives within a short period~\cite{monib2025microlearning,cronin2024microlearning}. Prior studies suggest that microlearning can support knowledge and skill acquisition better than some traditional training methods~\cite{de2019microlearning,cronin2024microlearning}. Related HCI work on embedded security training found that instruction delivered during an ongoing task was more effective than standard security notices~\cite{kumaraguru2007protecting}. This perspective also aligns with behavioral accounts in which timely prompts matter most when users have both the ability to act and a reason to do so~\cite{DBLP:conf/persuasive/Fogg09a}.

These studies suggest that comprehension support should be delivered where and when users make decisions. Interfaces can provide short explanations for unfamiliar terms, structure exposure through chunking and pacing, and include brief checks that encourage users to revisit relevant text when they miss key points.

\subsection{Usable Privacy}

Privacy policies and similar disclosures are difficult for most people to understand in one pass~\cite{wagner2023privacy,tang2021defining}. Usable privacy research has explored formats that make policy terms easier to find and compare, including structured presentations and standardized notice layouts~\cite{kelley2009nutrition,kelley2010standardizing}. Schaub et al.~\cite{schaub2015design} systematize these approaches by mapping a design space for privacy notices and identifying design dimensions that shape what users notice and how they interpret a notice. Prior HCI research has further shown that specific consent interface choices can substantially shape users' experiences and decisions. Habib et al.~\cite{habib2022okay} evaluated multiple cookie consent interface designs and found that design choices affected usability across several objectives. More recently, Xiao et al.~\cite{xiao2026designing} examined interface structures for privacy disclosures and found that presentation structure shaped attention and navigation but did not uniformly improve comprehension. Our study differs by incorporating comprehension assessment and, in some conditions, a second review into the consent process, and by examining initial comprehension, recovery after re-exposure, interaction cost, and consent behavior together.

Structured presentations can shape perceived privacy risk and willingness to proceed, but users may struggle when terminology is unfamiliar or when label information is misaligned with underlying practices~\cite{balash2024would,habib2022evaluating,DBLP:journals/popets/AliBKKA24}. Related work also shows that small interface elements can affect whether users recognize and correctly interpret available privacy choices~\cite{habib2021toggles}. Additional tools aim to reduce the burden of long disclosures by segmenting and summarizing policy text to help readers locate relevant clauses~\cite{woodring2024enhancing}.

We build on the usable privacy literature by treating structured presentation, pacing, explanation, and reassessment as forms of pedagogical friction. We examine not only initial comprehension, but also recovery after a second review, time cost, perceived burden, and the relationship between demonstrated comprehension and consent decisions. 

\subsection{Demonstrably Informed Consent}

In research ethics, informed consent is commonly described as involving information, comprehension, and voluntariness~\cite{hhs_ohrp_informed_consent_faq}. Several strands of work emphasize that these conditions can be supported or undermined by interface design~\cite{friedman2005informed,kim2017relative,millett2001cookies}. HCI research has also treated consent as a design concept. Im et al.~\cite{im2021yes} describe affirmative consent in interactive systems as being voluntary, informed, revertible, specific, and unburdensome, and use these properties to reason about the design of social computing systems.

The General Data Protection Regulation (GDPR) strengthens informational duties and places the burden of proving consent on controllers, but it primarily articulates consent in terms of awareness and demonstrability of the act of consenting, not verified understanding of specific terms~\cite{regulation2016regulation,european2020guidelines,frischmann2026defendingconsent}. This distinction motivates our empirical focus on privacy policy consent flows that aim to produce measurable evidence of comprehension.

Demonstrably informed consent places the responsibility for demonstrating comprehension onto the party seeking consent. Specifically, the party seeking consent should be able to produce evidence that a person both intended to consent and understood substantively important terms at the time consent was sought~\cite{frischmann2026defendingconsent}. A central practical issue is scope. Demonstrably informed consent does not require the same level of demonstrated comprehension for every clause, and prior work highlights that which terms are “substantive,” and how much detail must be understood, varies by context~\cite{mulligan2016privacy,kim2019consentability,frischmann2026defendingconsent}.

Our novelty lies in treating these mechanisms as parts of a consent process that links comprehension support, evidence of understanding, and interaction cost. The individual mechanisms we study, including highlighting, pacing, sectioning, brief explanations, comprehension checks, and re-exposure, draw on established interface design techniques. Our contribution is to organize them as a process design framing for demonstrably informed consent. %Pedagogical friction distinguishes support for initial comprehension from support during a second review, while treating interaction cost as part of the design outcome. 
We characterize this framing along three linked dimensions: how friction is presented; whether it supports initial review or re-review; and, the interaction costs it imposes.

The pedagogical friction framing differs from general friction in design by centering comprehension as the purpose of the intervention and by incorporating comprehension assessment into the consent process. It also differs from microlearning by linking brief instructional support to evidence of understanding. Unlike embedded instruction, which focuses on delivering support during an ongoing task, our framing treats both comprehension support and assessment as parts of the consent process.

\section{Method}
\label{sec:method}

\begin{figure}[t]
  \centering
  \includegraphics[width=\linewidth]{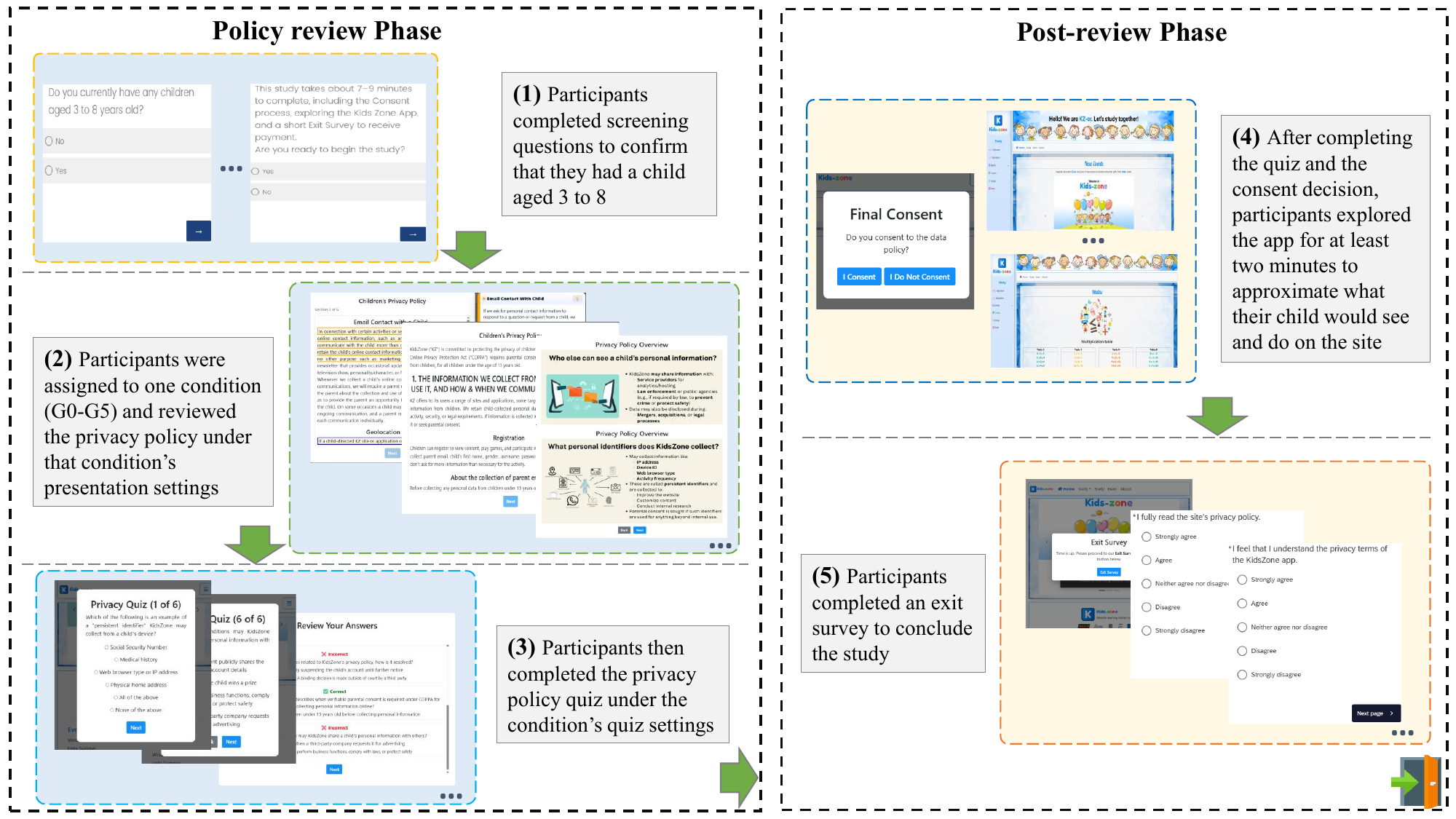}
  \caption{Study flow. Participants completed screening, reviewed the privacy policy under one of six assigned conditions, completed the comprehension quiz and consent decision, explored the learning app, and completed an exit survey.}
  \Description{Flow diagram with five stages. Participants complete screening, review the privacy policy under one of six assigned conditions, complete a comprehension quiz and consent decision, explore the learning app for two minutes, and complete an exit survey.}
  \label{fig:study_flow}
\end{figure}

We conducted a randomized experiment where participants completed a short privacy consent task as part of a multi-step study flow (see Figure~\ref{fig:study_flow}). We built a learning application for children by adapting public code from KidsZone~\cite{kidszone_Website}. The app includes various pages that teach children basic skills, such as the alphabet, counting, and simple math (see Appendix Figures~\ref{fig:kz_alphabet}-\ref{fig:kz_sub}).

Participants first completed a screening survey to confirm that they were the parent of a child between the ages of 3 and 8, the age range the learning app targets. Participants who did not meet these requirements were screened out. We then asked participants to imagine that their child wanted to use the KidsZone app and to decide whether to allow it. %take the role of a parent who is deciding whether their child can use the learning website. 
Before making this decision, they viewed the app's privacy policy. The policy explains what data the app may collect from children, how it may use and share that data, and what choices parents have, including consent and deletion requests (see Appendix Figure~\ref{fig:policy_fulltext} for the full terms). 

After viewing the policy, participants took a six-question quiz that matched substantive points in the policy (quiz questions provided in Appendix Table~\ref{tab:quiz_questions}). We used an 80\% score on the quiz as the basic threshold for demonstrated comprehension (a minimum of 5 of 6 questions answered correctly). In this study, demonstrated comprehension refers to immediate performance on the selected policy questions and does not imply broader or lasting understanding of the full policy. We treat this threshold as an operational study criterion, not as a legal or universal standard for informed consent. Whether this level of demonstrated comprehension is sufficient to support informed consent will depend on the context and the substantive terms at issue. Participants could not return to the policy pages during the quiz, so their answers relied on what they either learned from reviewing the policy or knew prior to the study.

After the quiz, participants made their consent decision under the particular settings of their experimental condition (§3.1). They then explored the app for a minimum of two minutes, followed by an exit survey about their experience with the consent process and their views of the app and its privacy policy.

To preserve study validity, we did not disclose in advance that the privacy policy consent flow itself was the focus of the experiment. The study protocol received prior approval from our Institutional Review Board (IRB).

\begin{figure}[t]
  \centering
  \includegraphics[width=\linewidth]{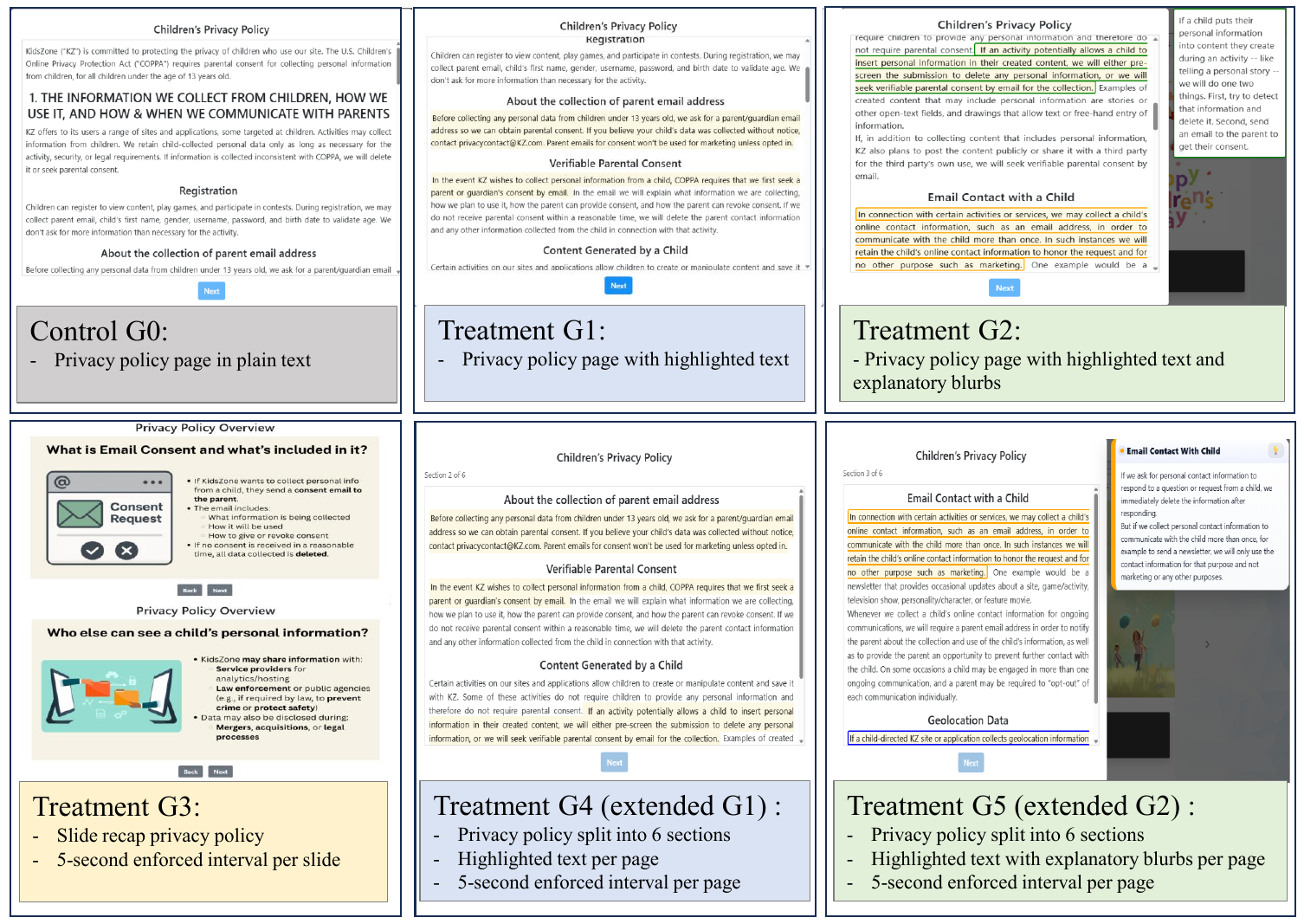}
  \caption{Six experimental conditions (G0-G5) for privacy policy review: (G0) plain text policy; (G1) highlighted policy; (G2) highlighted policy with explanatory blurbs; (G3) timed slide recap policy; (G4) sectioned highlighted policy with pacing; and (G5) sectioned highlighted policy with blurbs and pacing. Full-size examples of privacy policy review interfaces for the six conditions are provided in Appendix Figures~\ref{fig:g0_policy}-\ref{fig:g5_policy}.} 
  \Description{Six experimental conditions (G0-G5) for privacy policy review.}
  \label{Fig:groups}
\end{figure}

\subsection{Experimental Conditions}
\label{subsec:experimental_conditions}

After enrollment and screening, participants were randomly assigned to one of six experimental conditions (G0-G5) for privacy policy review. As shown in Figure~\ref{Fig:groups} and summarized in Table~\ref{table:design_matrix}, the conditions differed in how the policy was presented and how the quiz was administered. Notably, participants in all six conditions received the same, full privacy policy text. The text was presented in a scrollable panel, and participants could not advance until they had scrolled to the end of the text on each page. This ensured that every participant was exposed to the complete policy, though it does not establish that they read it. What varied were the additional forms of pedagogical friction.

\noindent \textbf{Control group (G0): Plain text policy.}
Participants viewed a standard privacy policy page in plain text and then took the quiz once. After the quiz, they were shown which questions they answered incorrectly and the correct answers. Participants who scored at least 80\% could choose whether to provide consent to the privacy policy terms or to decline consent. Participants who scored below 80\% were still allowed to choose whether to consent; we use this condition to measure consent without demonstrated comprehension.

\noindent \textbf{Treatment group 1 (G1): Highlighted policy.}
Participants viewed the same policy page, but key policy statements were highlighted. The highlighted text covered the information needed to answer the quiz questions. Apart from highlighting, all other procedures were identical to G0, including that consent was not gated on the quiz threshold.

\noindent \textbf{Treatment group 2 (G2): Highlighted policy with explanatory blurbs.}
Participants viewed the highlighted policy and also saw short explanatory blurbs next to the highlighted statements. The blurbs were designed to explain the highlighted terms in simpler language. Participants first took the quiz once. If they scored below 80\%, they were shown which questions they answered incorrectly, but not the correct answers. They then reviewed the policy again and retook the quiz. After the final quiz attempt, participants who scored at least 80\% could choose whether to provide consent to the privacy policy terms or to decline consent. Participants who scored below 80\% could not provide consent.

\noindent \textbf{Treatment group 3 (G3): Timed slide recap policy.}
Participants first viewed the standard privacy policy page in plain text and then viewed a recap of the same terms in slide format. Each slide posed a question about one key term and answered it with icons or illustrations and a short, bulleted explanation in simplified language (7 slides total; see Appendix Figure~\ref{fig:g3_policy}). %The slide format was designed to visually recap the key terms in a Q\&A format. 
A 5-second minimum time was enforced on each slide before participants could continue. Participants then took the quiz once. If they scored below the 80\% threshold, they reviewed the slides again---without the enforced minimum and without returning to the full policy text---and retook the quiz. Quiz and gating procedures were otherwise identical to G2.

\noindent \textbf{Treatment group 4 (G4): Extended G1.}
This condition extended G1 by splitting the highlighted policy into six sections, with one section per page. A 5 second minimum time was enforced on each page before participants could continue. Apart from these changes, all other procedures were identical to G1, and consent was not gated on the quiz threshold.

\noindent \textbf{Treatment group 5 (G5): Extended G2.}
This condition extended G2 by splitting the highlighted policy into six sections, with one section per page. Each page included both highlighted statements and explanatory blurbs. A 5 second minimum time was enforced on each page before participants could continue. Apart from these changes, all other procedures were identical to G2.

\begin{table}[t]
\caption{Design matrix for the six experimental conditions. All conditions
presented the same, full privacy policy text; conditions differ in the
pedagogical friction added to policy review and in the quiz procedure.
\cmark~indicates the feature was present. Pacing refers to an enforced
5-second minimum per page or slide. In G3, the enforced minimum applied to the seven recap slides and not to the policy text. Retake (gated) indicates that
participants who did not meet the 80\% threshold on the first attempt
reviewed the policy material again and retook the quiz, and that participants who did
not meet the threshold on the final attempt could not provide consent; in
all other conditions the quiz was taken once and consent was available
regardless of the score.}
\label{tab:design_matrix}
\small
\begin{adjustbox}{max width=\linewidth}
\centering
\begin{tabular}{@{}llccccccc@{}}
\toprule
& & \multicolumn{6}{c}{Policy review} & Quiz \\
\cmidrule(lr){3-8}\cmidrule(lr){9-9}
Group & Short name & Plain-text policy & Highlight & Blurbs & Slide recap & Sect. & Pacing & Retake (gated) \\
\midrule
G0 & Plain-text policy            & \cmark &         &         &         &         &         &         \\
G1 & Highlighted           & \cmark & \cmark  &         &         &         &         &         \\
G4 & Sectioned + paced     & \cmark & \cmark  &         &         & \cmark  & \cmark  &         \\
G2 & Highlighted + blurbs  & \cmark & \cmark  & \cmark  &         &         &         & \cmark  \\
G5 & Sectioned + blurbs    & \cmark & \cmark  & \cmark  &         & \cmark  & \cmark  & \cmark  \\
G3 & Timed slide recap     & \cmark &         &         & \cmark  &         & \cmark & \cmark  \\
\bottomrule
\end{tabular}
\end{adjustbox}
\label{table:design_matrix}
\end{table}

We note that G3 differs structurally from the other conditions. Rather than modifying how the policy itself was presented, it added a second, simplified presentation of the key terms after the full policy. G3 participants therefore encountered the key terms twice before the first quiz. We also note that retake and gating covary by design. The second attempt provides the opportunity to meet the threshold that gating requires. We therefore do not interpret differences between gated and ungated conditions as effects of gating alone (§4.5).

Our design enables comparisons across conditions with different forms and levels of pedagogical friction and associated costs. The conditions were designed to support incremental comparisons, with several conditions adding forms of pedagogical friction relative to earlier conditions. We interpret the results primarily at the condition level rather than attributing differences to any single element.

%\subsection{Study Design}
%\label{subsec:procedure}

%Our study consisted of a friction-based policy review phase followed by a post-review phase.

%\noindent\textbf{Friction-based policy review phase}.

%\noindent   \textbf{(1)} Participants completed screening questions to confirm that they had a child and were eligible to take part in the study.

%\noindent   \textbf{(2)} Participants were assigned to one condition (G0-G5) and reviewed the privacy policy under that condition’s presentation settings.

%\noindent \textbf{(3)} Participants then completed the privacy policy quiz under the condition’s quiz settings.

%\noindent \textbf{Post-review phase}.

%\noindent \textbf{(4)} After completing the quiz and the consent decision, participants explored the website for two minutes to mimic a brief parent check of the site content.

%\noindent   \textbf{(5)} Participants then completed an exit survey. The survey asked about their experience with the consent process and their views of the website and its privacy policy, and then the study ended.

\subsection{Participants and Exclusions}
Participants were recruited via Positly.\footnote{\url{https://www.positly.com}} Eligibility required that participants be at least 18 years old and have at least one child between the ages of 3 and 8. Sessions lasted approximately 8 minutes, and participants were paid \$18.75 per hour, prorated to their completion time. Of the 1207 participants who began the study, 914 did not complete it and were excluded. Most incomplete sessions ended near the beginning of the study and did not contribute usable study responses. The final analytic dataset comprised 293 participants, randomly assigned to the six conditions and distributed as follows: G0 (46), G1 (47), G2 (48), G3 (48), G4 (49), and G5 (55).

We conducted an a priori power analysis in G\textasteriskcentered Power for the omnibus comparison of first attempt mean quiz accuracy across six groups. We assumed a medium effect size of $f=0.25$, $\alpha=0.05$, and target power of $0.9$, which indicated a required total sample size of 270. Our final analytic sample (293) exceeded this target. The analysis did not separately power threshold attainment or pairwise condition comparisons.

\subsection{Demographics}

Participant demographics provided by Positly are summarized in Appendix Table~\ref{tab:demographics}. The sample was 54.6\% female and 45.4\% male, and most participants were 31 to 40 years old (50.7\%). Education levels were high school or below (30.9\%) and bachelor's degree (30.9\%), followed by graduate or professional degree (20.3\%). Household income was broadly distributed, with most participants reporting \$60{,}000 to \$149{,}999 per year (56.0\%).

We used chi squared tests to determine whether demographics differed across the six groups. We found no significant differences across groups for age category, gender, or education (all $p \geq 0.05$), which is consistent with successful random assignment.

\subsection{Instrumentation and Measures}
\label{subsec:instrumentation_measures}
We recorded policy review time, quiz completion time, timestamps, and quiz responses for each participant. These logs let us measure engagement and objective time costs of friction. They also allowed us to quantify policy comprehension via quiz accuracy. We collected exit survey responses to capture participants' perceived comprehension and their experience with the consent process, including self-reported burden and whether the time felt worthwhile.

\subsubsection{Interaction telemetry.} 
Telemetry variables describe how participants engaged with the consent flow. They are not the main outcomes, but they help interpret differences across conditions. We measure:

\begin{itemize}
\item \textbf{Policy review duration.} 
Time spent viewing the policy content (reported in seconds).

\item \textbf{Quiz completion duration.}
Time spent taking the quiz (reported in seconds).
\end{itemize}

\subsubsection{Policy comprehension measures.}
Participants completed a six-question comprehension quiz tied to specific policy statements (see Appendix Table~\ref{tab:quiz_questions}). Quiz accuracy is the number of correct answers divided by six.

\begin{itemize}
\item \textbf{First quiz accuracy.} Mean accuracy on the first quiz attempt within each condition.

\item \textbf{Retake accuracy.} Second-attempt accuracy among participants who went on to retake the quiz (G2, G3, and G5). This isolates the subgroup that did not meet the threshold on the first attempt.

\item \textbf{Retake gain.} The change in accuracy from the first to the second attempt for retake participants (second accuracy minus first accuracy). We report the average gain by condition.

\item \textbf{Retake improvement rate.} Among participants who took the second quiz, the proportion whose second score was higher than their first score.

\item \textbf{Threshold attainment.} The proportion of participants who scored at least 80\%, which we treat as demonstrated comprehension in all conditions.

\end{itemize}

\subsubsection{Exit survey measures.}
The exit survey used 5-point Likert items (strongly agree to strongly disagree) to measure participants' perceived comprehension, perceived helpfulness of the interface features, self-reported burden of the consent process, and whether the time spent in the consent process felt worthwhile. Most exit survey items were positively framed, but item~8 was negatively framed and asked whether the consent process felt like a burden. We use these items to summarize participant experience and to support the case-based analysis reported in \S~\ref{sec:results}. We operationalize observed costs through objective time-on-task measures and use self-reported burden and whether the time felt worthwhile as supporting indicators of participants' reactions to the study. Self-reports are not direct measures of cognitive load or mental effort.

\begin{table}[t]
\centering
\caption{Quiz performance by group. Threshold attainment was reported with Wilson 95\% confidence interval (CI). ``--'' denotes that the group had no retake design.}
\label{tab:rq1_accuracy_bygroup}

\small
\begin{adjustbox}{max width=\linewidth}
\begin{tabular}{
    >{\centering\arraybackslash}p{0.8cm}
    >{\centering\arraybackslash}p{1.0cm}
    >{\centering\arraybackslash}p{1.25cm}
    >{\centering\arraybackslash}p{2.6cm}
    >{\centering\arraybackslash}p{1.2cm}
    >{\centering\arraybackslash}p{1.0cm}
    >{\centering\arraybackslash}p{1.25cm}
    >{\centering\arraybackslash}p{2.7cm}}
\toprule
\multirow{2}{*}{\textbf{Group}} &
\multicolumn{2}{c}{\textbf{\makecell{Quiz participants\\$N$}}} &
\multicolumn{1}{c}{\textbf{\makecell{Reached $\ge$80\% on quiz\\\% (Count) [95\% CI]}}} &
\multicolumn{1}{c}{\textbf{\makecell{Quiz mean\\accuracy}}} &
\multicolumn{2}{c}{\textbf{\makecell{Retakers' quiz\\mean accuracy}}} &
\multicolumn{1}{c}{\textbf{\makecell{Reached $\ge$80\% on quiz\\\% (Count) [95\% CI]}}} \\
\cmidrule(lr){2-3}
\cmidrule(lr){4-4}
\cmidrule(lr){5-5}
\cmidrule(lr){6-7}
\cmidrule(lr){8-8}
& First attempt & Retake subgroup & First attempt
& First attempt
& First attempt & Retake subgroup & Retake subgroup \\
\midrule
G0 & 46 & -- & 17.4 (8) [9.1--30.7]  & 0.52 & --   & --   & -- \\
G1 & 47 & -- & 19.1 (9) [10.4--32.5] & 0.54 & --   & --   & -- \\
G2 & 48 & 38 & 20.8 (10) [11.7--34.3] & 0.54 & 0.42 & 0.56 & 18.4 (7) [9.2--33.4] \\
G3 & 48 & 28 & 41.7 (20) [28.8--55.7] & 0.60 & 0.42 & 0.55 & 21.4 (6) [10.2--39.5] \\
G4 & 49 & -- & 30.6 (15) [19.5--44.5] & 0.63 & --   & --   & -- \\
G5 & 55 & 47 & 14.5 (8) [7.6--26.2]  & 0.52 & 0.46 & 0.63 & 34.0 (16) [22.2--48.3] \\
\bottomrule
\end{tabular}%
\end{adjustbox}
\end{table}

\section{Results}
\label{sec:results}
%Our results are organized in two stages. 
%Results are organized by research question.  We first report inferential comparisons for first-attempt outcomes that are defined across all six conditions. We then report descriptive analyses for outcomes among retakers and consent patterns under gated versus ungated flows.\\

We organize the results around our two research questions and the paper’s central concern with the relationship between comprehension and consent. For RQ1, we first compare initial comprehension across all six conditions, then examine second-attempt outcomes in the three conditions that included a retake. For RQ2, we examine the time costs of the interventions and participants’ experiences of those costs. We then examine how demonstrated comprehension aligned with consent decisions, and conclude with exploratory analyses of participant-level and demographic patterns.

\noindent \textbf{RQ1}: How does demonstrated comprehension of key terms vary across privacy policy review designs with different forms of pedagogical friction?

\subsection{RQ1: Quiz performance by condition}
\label{subsec:rq1_quiz_by_condition}

\begin{figure}[t]
  \centering
  \includegraphics[width=0.65\linewidth]{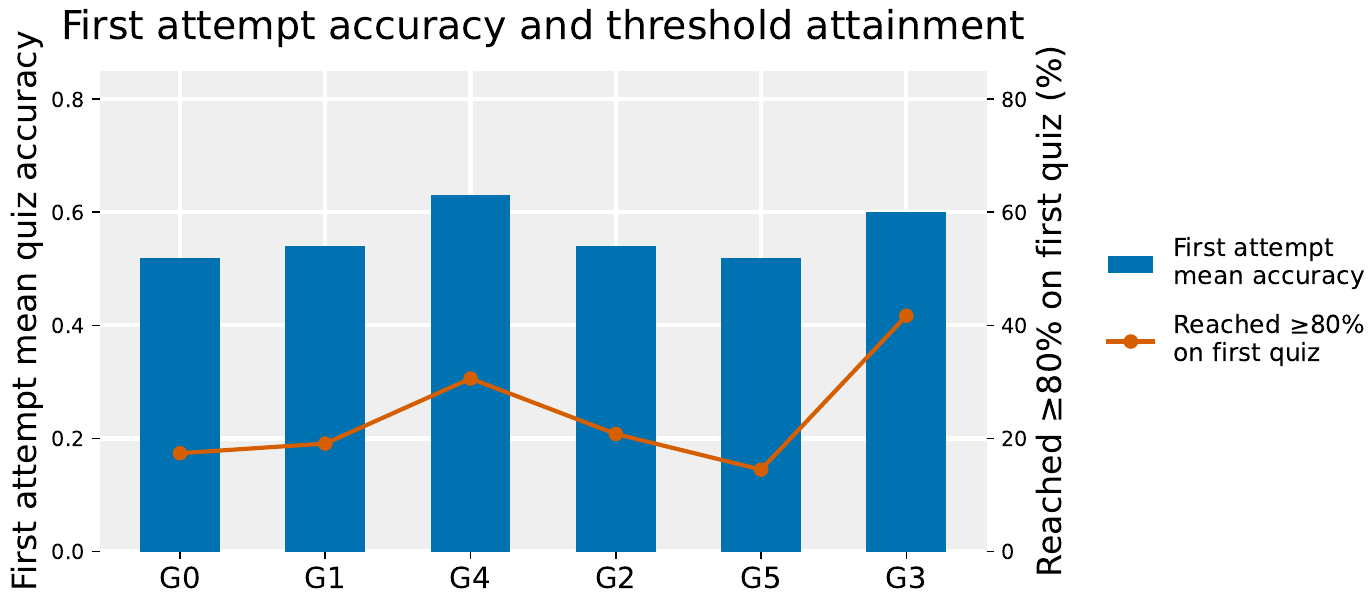}
  \caption{First-attempt quiz accuracy and threshold attainment by condition. G3 and G4 had the highest rates of reaching the 80\% comprehension threshold. Groups are ordered to reflect design lineage.}
  \label{fig:first_accuracy}
  \Description{First-attempt quiz accuracy (bars) and threshold attainment (line)}
\end{figure}

We first compare performance on the initial privacy policy quiz across conditions. We also report the proportion of participants who met the comprehension threshold on the first attempt (details are provided in Table~\ref{tab:rq1_accuracy_bygroup}).

First-attempt mean accuracy in the control condition G0 (plain text) was 0.52 (Figure~\ref{fig:first_accuracy}), comparable to G5 (paced, sectioned with blurbs; 0.52), and slightly lower than G1 (highlighted; 0.54) and G2 (highlighted with blurbs; 0.54). The highest first-attempt mean accuracies were in G3 (timed slide recap; 0.60) and G4 (paced, sectioned highlighting; 0.63).

%A clearer separation appears when we consider the first-attempt threshold outcome. 
The conditions were more differentiated on first-attempt threshold attainment. Threshold attainment was 17.4\% in G0, 41.7\% in G3, and 30.6\% in G4. We ran a one-way ANOVA with condition as a between-subjects factor to test for differences in first-attempt mean quiz accuracy ($F(5,287)=1.77$, $p=.12$). Differences in average accuracy were modest relative to within-condition variability. In contrast, first-attempt threshold attainment differed by condition overall ($\chi^2(5,293)=14.1$, $p=.015$). In Holm-adjusted pairwise comparisons, G3 versus G5 was the only contrast that remained statistically reliable ($p=.03$). 

The timed slide recap condition G3 had the highest first-attempt threshold attainment, followed by the paced, sectioned highlighting condition G4. G3 should be interpreted as the complete condition, in which participants viewed the full plain-text policy and then encountered a simplified, illustrated recap of the key terms before the quiz. Its advantage may reflect this repeated exposure in a second format rather than the slides or pacing in isolation. %which combined the plain policy with an additional timed slide recap before the quiz, not as an isolated effect of slides or pacing.

Question accuracy shows that difficulty was not evenly distributed across quiz questions (see Appendix Table~\ref{tab:quiz_questions}). On the first attempt, Q4 was the hardest question overall (32.4\% correct), followed by Q5 (42.0\%), while Q2 was the easiest (78.2\%). This pattern was broadly consistent across conditions, with Q4 appearing as the hardest question in every group. Among retakers, Q4 remained the hardest question on the second attempt (40.7\% correct) and was also the hardest to fix. Among participants who initially answered Q4 incorrectly, 68.5\% still answered it incorrectly on the second attempt.

\subsubsection{Additional analyses of first attempt comprehension}

We conducted two additional analyses to examine the consistency of the first-attempt patterns. First, we fit a logistic model using generalized estimating equations across individual quiz responses, accounting for quiz question and clustering responses within participants. G3 and G4 again showed the largest positive estimates relative to G0 (G3: OR=1.44, nominal 95\% CI [0.90, 2.30]; G4: OR=1.68, nominal 95\% CI [1.07, 2.64]). After Holm correction, neither comparison remained statistically reliable.

Second, because our primary 80\% comprehension threshold corresponds to answering at least 5 of the 6 questions correctly, we examined whether the patterns changed under nearby alternative thresholds. We compared attainment at 4/6, 5/6, and 6/6 correct answers. G3 and G4 had the two highest observed attainment rates under both the 4/6 and 5/6 criteria. When requiring all six answers to be correct, the condition pattern changed and the overall difference was not statistically reliable. These results suggest that the higher observed attainment in G3 and G4 is not unique to the 80\% cutoff.

\subsection{RQ1: Retake outcomes for conditions with a second attempt}
\label{subsec:rq1_retake}

\subsubsection{Retake Gains on Quiz Performance}
\begin{figure}[t]
  \centering
  \includegraphics[width=0.7\linewidth]{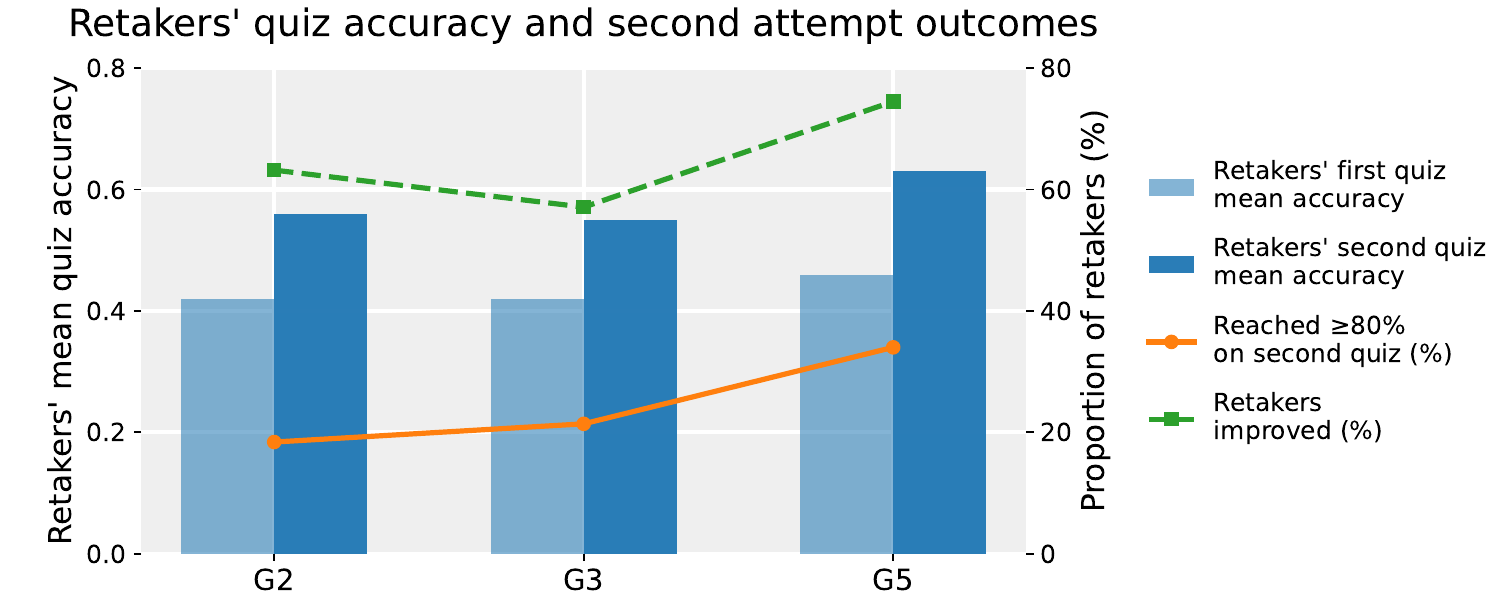}
  \caption{Retakers' mean accuracy, second-attempt threshold attainment, and improvement proportion by condition. G5 (paced, sectioned with blurbs) has the highest observed second-attempt threshold attainment and improvement.}
  \label{fig:second_accuracy}
  \Description{Retakers' mean quiz accuracy (bars), second quiz threshold attainment (solid line), and the proportion of retakers whose score increased (dashed line), by condition.}
\end{figure}

Only G2 (highlighted with blurbs), G3 (timed slide recap), and G5 (paced, sectioned with blurbs) included a second quiz attempt. Participants who scored below the 80\% threshold reviewed the policy material again and then retook the quiz. We report retake accuracy, retake gain, and the proportion of retakers whose score increased (Figure~\ref{fig:second_accuracy}). We treat differences among retake conditions as descriptive and use them to characterize observed recovery patterns.

Scores increased from the first to the second attempt across all three retake conditions, with G5 showing the strongest observed recovery pattern. In G2, 38 participants took the second attempt, with a mean accuracy gain of +0.14, and 63.2\% improved. In G3, 28 participants retook the quiz, with a mean gain of +0.13, and 57.1\% improved. In G5, 47 participants retook the quiz, with a mean gain of +0.17, and 74.5\% improved. Overall, 75 of 113 retakers (66.4\%) improved their score on the second attempt.

\subsubsection{Threshold Attainment after Retake}
A stricter measure of retake outcome is whether retakers demonstrated comprehension on the second attempt by reaching the 80\% threshold. G5 (paced, sectioned with blurbs) showed the highest recovery among retakers: 34.0\% reached the 80\% threshold, compared to 18.4\% in G2 (highlighted with blurbs) and 21.4\% in G3 (timed slide recap). Reaching 80\% is the operational threshold we set for demonstrated comprehension in this study. G3 retakers reviewed only the slide recap, without the enforced per-slide minimum applied on the first attempt (median 21s; Appendix Table 7), so its retake imposed considerably less exposure than the retakes in G5. These results are consistent with stronger recovery in G5 among participants who did not meet the threshold on their first attempt.%initially scored below the threshold.

\subsubsection{Error Correction Dynamics During Retake}
\begin{figure}[t]
  \centering
  \includegraphics[width=0.68\linewidth]{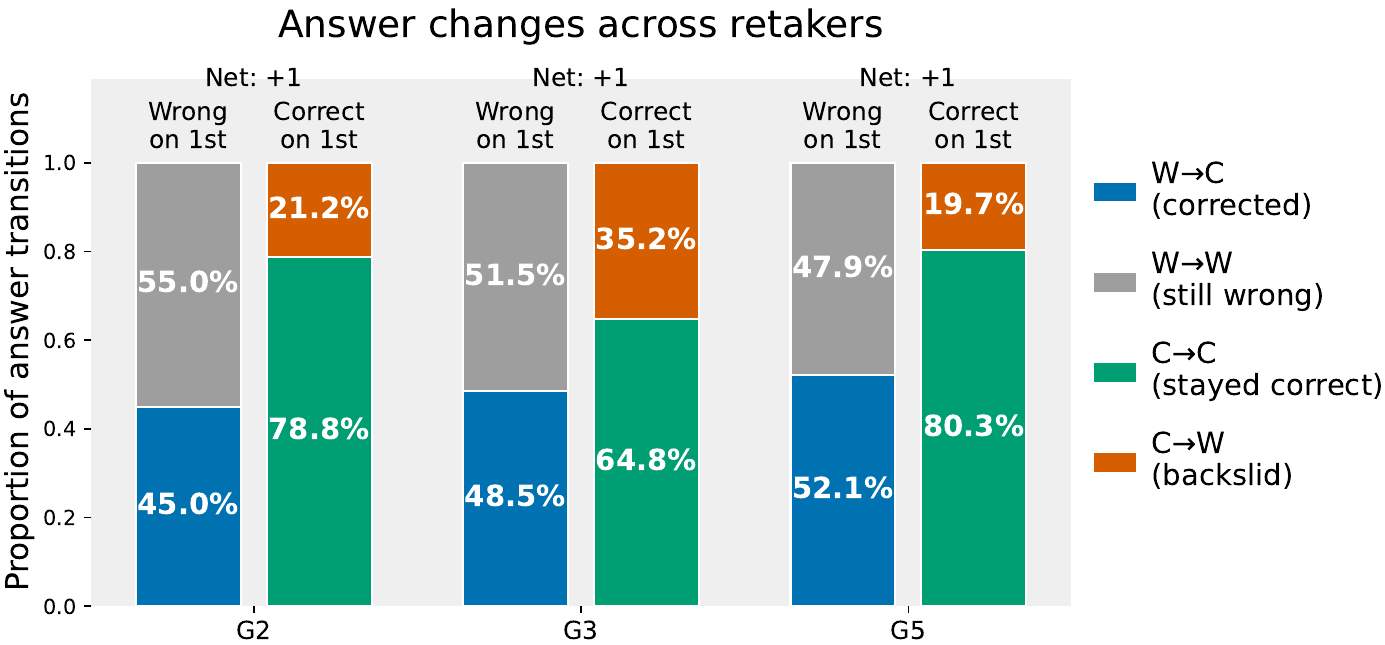}
  \caption{Answer changes across retakers, conditional on first-attempt correctness. Net is the median per-retaker (wrong$\rightarrow$correct) minus (correct$\rightarrow$wrong), positive values indicate more fixes than new mistakes.}
  \label{fig:answer_changes}
  \Description{Answer changes across retakers, conditional on first-attempt correctness (left: first wrong; right: first correct). \emph{Net} is per-retaker (wrong$\rightarrow$correct) minus (correct$\rightarrow$wrong), where positive values indicate more fixes than new mistakes.}
\end{figure}

We analyze answer transitions across the six quiz questions. We count (1) \emph{corrections} (wrong$\rightarrow$correct), (2) \emph{backslides} (correct$\rightarrow$wrong), (3) \emph{persistent errors} (wrong$\rightarrow$wrong), and (4) \emph{stable correct} (correct$\rightarrow$correct) (Figure~\ref{fig:answer_changes}).

G5 (paced, sectioned with blurbs) had the highest correction rate and the lowest observed backsliding rate. Among questions answered incorrectly on the first attempt, retakers corrected 52.1\% in G5, compared to 45.0\% in G2 (highlighted with blurbs) and 48.5\% in G3 (timed slide recap). Among questions answered correctly on the first attempt, backsliding was lowest in G5 (19.7\%). Persistent errors were correspondingly lowest in G5, where 47.9\% of initially incorrect questions remained incorrect. Retakers typically fixed about one more question than they newly missed on the second attempt (median net gain was $+1$ question).

These transition patterns are consistent with the larger retake gains and higher second-attempt threshold attainment observed in G5, the only retake condition that combines sectioned policy pages with an enforced 5-second interval per section.\\

\noindent \textbf{RQ2}: What costs accompany frictional interventions and how are they tolerated by users?

\subsection{RQ2: Time cost}
\label{subsec:time_cost}

We analyze first- and second-attempt policy review time, quiz completion time, and their total time. Second-attempt times apply only in G2  (highlighted with blurbs), G3 (timed slide recap), and G5 (paced, sectioned with blurbs). Time distributions are right-skewed, so we summarize time (seconds) using medians and interquartile ranges. We compare first-attempt policy review and quiz completion times across conditions using Kruskal-Wallis tests and report epsilon squared effect sizes.

First-attempt policy review time differed across conditions
($p<.001$, $\epsilon^2=.42$), with the paced and structured designs showing longer review times (e.g., median 55s in G4 vs. 10s in G0; Appendix Table~\ref{tab:time_cost_all}). In G3, viewing of the plain-text policy and of the slide recap were logged as a single duration, so the 70s median covers both and is not directly comparable to the policy-only times in other conditions. Quiz completion time also differed across conditions ($p=.006$, $\epsilon^2=.07$), although the effect was substantially smaller.

\subsection{RQ2: Participants’ experience of friction}
\label{subsec:friction_tolerance}

We use the exit survey to characterize participants’ experiences of the interventions, particularly perceived burden and whether the time spent felt worthwhile (see Table~\ref{tab:exit_survey} for survey questions). Across all answered Likert items, 75\% of responses were Strongly agree or Agree (Table~\ref{tab:exit_survey_summary}). Few respondents gave the same rating to every item (overall 4.8\%), suggesting limited straightlining.

Self-reported comprehension was high across conditions (overall endorsement 85.4\%), with modest variation by group. Burden endorsement was lowest in lighter friction conditions (G0 and G1) and higher in conditions with more enforced exposure and retakes (e.g., G2, G5), corresponding to a modest association between condition and burden endorsement (Cramer's $V\approx.22$). We treat these responses as supporting context and interpret them alongside observed behavior and quiz outcomes in illustrative cases (\S~\ref{subsec:case_study}).

\subsection{Consent Decisions}
\label{subsec:consent_decision}

%We define consent rate as the fraction of participants who selected consent among those who demonstrated comprehension on the quiz ($\ge$80\%). 

\begin{figure}[t]
  \centering
  \includegraphics[width=0.65\linewidth]{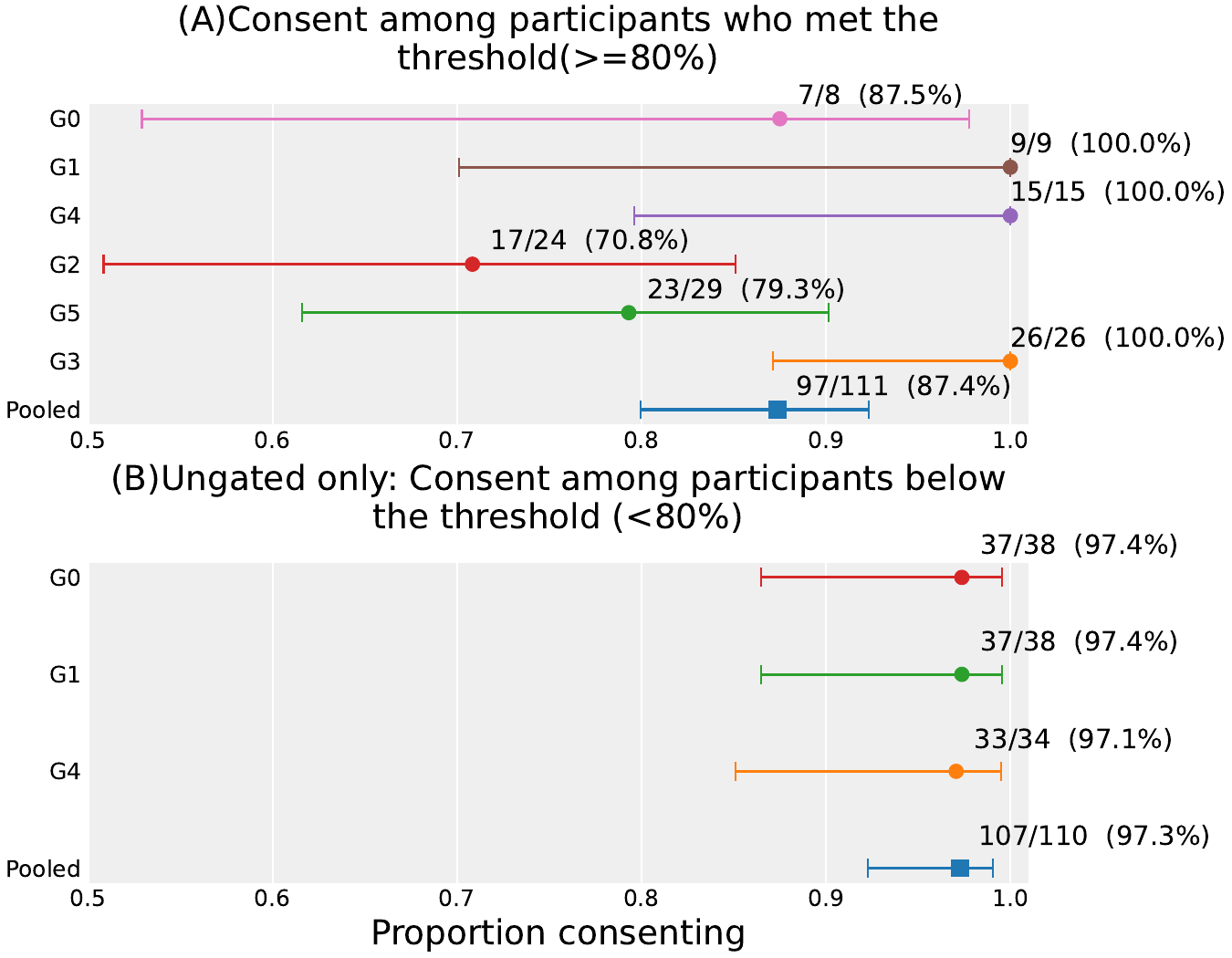}
  \caption{Consent rates by condition, stratified by demonstrated comprehension on the quiz. (A) shows the proportion who selected consent among participants who met the comprehension threshold ($\geq$80\%). (B) shows, for ungated conditions only (G0, G1, G4), the proportion who selected consent among participants who did not meet the threshold ($<80\%$). Text labels of the form $x/y$ indicate the number who selected consent out of the number in that subset, with the corresponding percentage. Points show observed consent proportions with Wilson 95\% CIs.}
  \label{fig:consent_rate}
  \Description{Consent rates by condition, stratified by whether participants met the comprehension threshold ($\ge$80\%). The below panel includes only ungated conditions (G0, G1, and G4). Points show observed consent proportions with Wilson 95\% CIs.}
\end{figure}

We next examine the relationship between demonstrated comprehension and consent. The clearest comparison comes from the ungated conditions (G0, G1, and G4), where participants could consent regardless of their quiz performance. In these conditions, consent was nearly identical among participants who met the comprehension threshold (96.9\%) and those who did not (97.3\%) (Figure \ref{fig:consent_rate}). Thus, recorded agreement remained high regardless of whether participants demonstrated comprehension of the policy terms.

We also report consent among participants who met the comprehension threshold across all six conditions. For G0, G1, and G4, threshold status is based on the first quiz attempt; for the retake conditions G2, G3, and G5, it is based on the final quiz attempt completed. Across these participants, the pooled consent rate was 87.4\%, ranging from 70.8\% in G2 to 100\% in G1, G3, and G4 (Figure \ref{fig:consent_rate}).

These results should not be interpreted as evidence that gating changes willingness to consent. Participants who remained below the threshold in G2, G3, and G5 were not permitted to consent, so consent rates cannot be directly compared between gated and ungated conditions. Rather, the ungated conditions reveal the central pattern. In particular, agreement was not meaningfully differentiated by demonstrated comprehension.

\begin{table}[t]
\centering
\caption{Exit Survey Questions.}
\label{tab:exit_survey}
\small
\begin{adjustbox}{max width=\linewidth}

\begin{tabular}{
    >{\centering\arraybackslash}p{1.0cm}
    >{\raggedright\arraybackslash}p{8.8cm}}
\toprule
\textbf{Item} & {\centering\textbf{Question}} \\
\midrule
1 & I would let my children use the Kids Zone app. \\
2 & I fully read the site's privacy policy. \\
3 & The highlighting of parts of the privacy policy helped draw my attention to important terms. \\
4 & The explanatory blurbs which appeared alongside the highlighted text helped me to understand the important terms. \\
5 & The visual summaries (slides) which appeared after the privacy policy helped me to understand the important terms. \\
6 & I feel that I understand the privacy terms of the KidsZone app. \\
7 & I was satisfied with the length of the comprehension quiz. \\
8 & The time it took me to complete the consent process felt like a burden. \\
9 & The time it took me to complete the consent process was worthwhile. \\
\bottomrule
\end{tabular}
\end{adjustbox}
\end{table}

\subsection{Exploratory relationships between self-report and observed behavior}
\label{subsec:case_study}

%We use a small set of illustrative comparisons to connect exit survey responses with observed behavior (e.g., policy reading time) and quiz outcomes.
We conducted exploratory analyses to examine how participants’ self-reported experiences aligned with observed behavior and quiz performance. These analyses are descriptive and are intended to provide context for the aggregate results above.

\subsubsection{Self-reported policy reading and observed behavior}
%Participants who said they did not fully read the privacy policy also tended to spend less time on the policy and score lower on the first quiz attempt. 
Thirteen participants disagreed or strongly disagreed that they had fully read the privacy policy (exit survey item~2, Table~\ref{tab:exit_survey}). These participants generally spent less time reviewing the policy and had lower first-attempt quiz scores than their condition averages. Most of these cases coming from conditions with lighter friction, particularly G0 (plain text) and G1 (highlighted). In G0, participants who reported not fully reading the policy spent an average of 6.5 seconds reviewing it (vs.\ the group mean of 18.6 seconds) and their first-attempt accuracy was 0.42 (vs.\ the group mean of 0.52). A similar pattern appeared in G2 (highlighted with blurbs), where participants who reported not fully reading the policy also spent less time reviewing it, although their accuracy was closer to the group average. These descriptive patterns suggest that self-reported policy reading was broadly consistent with observed review behavior and, in some cases, quiz performance. %These cases align with the quantitative results, showing that when the interface makes it easy to skim, some participants do so, and quiz performance reflects that.

\subsubsection{Self-reported comprehension and retake performance}
%Participants who reported lower comprehension also showed improvement after retaking the quiz. 
Among participants in G2 (highlighted with blurbs) and G5 (paced, sectioned with blurbs) who reported low perceived comprehension (disagree or strongly disagree on item~6), first-attempt accuracy was lower than the average across those conditions (0.44 vs.\ 0.53). Their second-attempt accuracy increased to 0.57, approaching the mean among retakers in those conditions (0.59). A similar pattern appeared among participants who disagreed or strongly disagreed that the explanatory blurbs helped them understand the important terms (item 4): all required a retake, and their mean accuracy increased from 0.45 to 0.56. These comparisons suggest that low self-reported comprehension or perceived helpfulness did not preclude improvement on the second attempt. %provide additional context for the observed gains among retakers, including participants who reported that the explanatory content was not helpful.

%\subsubsection{Negative experiences and improved quiz performance}
%Improved second-attempt outcomes can occur even when participants report a negative experience. In G5, three participants disagreed or strongly disagreed with all relevant experience questions (items~2-4 and~6-8), yet their second-attempt scores were either unchanged or substantially higher (e.g., 0.14 to 0.71; 0.43 to 0.71). These cases illustrate that improved quiz performance and negative subjective reactions can occur together.

Taken together, these exploratory analyses show that participants’ subjective assessments of the consent process did not always align with their observed behavior or quiz performance. This reinforces the value of considering self-report, interaction behavior, and demonstrated comprehension as distinct outcomes. %These cases provide additional context for two broader patterns in the quantitative results: (1) G3 showed the highest observed first-attempt threshold attainment, and (2) many participants who initially struggled improved after a second policy review and quiz attempt.

\begin{table}[t]
\centering
\small
\caption{Exit survey items on consent experience.}
\label{tab:exit_survey_summary}
\begin{tabular}{
    l
    >{\raggedleft\arraybackslash}p{0.14\linewidth}
    >{\raggedleft\arraybackslash}p{0.18\linewidth}
    >{\raggedleft\arraybackslash}p{0.14\linewidth}}
\toprule
\multicolumn{4}{l}{\textbf{Response distribution}} \\
\midrule
\multicolumn{3}{l}{Response option} & Proportion \\
\midrule
\multicolumn{3}{l}{Strongly agree} & 37.6\% \\
\multicolumn{3}{l}{Agree} & 37.4\% \\
\multicolumn{3}{l}{Neither agree nor disagree} & 12.1\% \\
\multicolumn{3}{l}{Disagree} & 8.6\% \\
\multicolumn{3}{l}{Strongly disagree} & 4.3\% \\
\midrule
Group & Straightlining & Comprehension & Burden \\
\midrule
G0 & 0 & 81.5\% & 17.4\% \\
G1 & 8.5\% & 81.9\% & 19.1\% \\
G2 & 6.1\% & 86.7\% & 40.8\% \\
G3 & 4.2\% & 88.5\% & 31.2\% \\
G4 & 0 & 87.8\% & 24.5\% \\
G5 & 9.1\% & 85.5\% & 43.6\% \\
Overall & 4.8\% & 85.4\% & -- \\
\bottomrule
\end{tabular}

\footnotesize
\textit{Notes.} For the response distribution, the reported proportions pool all answered Likert items across participants. Straightlining is calculated per participant. Comprehension endorsement is based on item~6 within each group, and burden endorsement is based on item~8.
\end{table}

\subsection{Demographic Patterns}
\label{subsec:demo_outcomes}
We conducted demographic analysis to assess whether quiz outcomes varied across age and education groups (Appendix Table~\ref{tab:demographics_outcomes}). Across age groups, first-attempt mean accuracy was similar for participants aged 19-40 (0.54) and was higher for ages 41-50 (0.60) and 51-64 (0.64). First-attempt threshold attainment showed a similar pattern, increasing from 25.0\% (ages 19-30) to 33.3\% (51-64). Second-attempt mean accuracy ranged from 0.51 to 0.66 across the younger and middle-age groups.

Across education groups, first-attempt mean accuracy was similar for high school, associate degree, and bachelor's degree (0.50-0.52), and higher for graduate degree (0.62). First-attempt threshold attainment followed the same direction, rising from 11.8\% (high school) to 32.1\% (graduate degree). 

First-attempt mean accuracy and threshold attainment were descriptively higher in the 41-64 age groups than in the younger group. Second-attempt threshold attainment was descriptively higher in the high school and associate degree groups than in the bachelor's and graduate degree groups. We treat these subgroup patterns as descriptive because the study was not powered for demographic subgroup comparisons.

\section{Discussion}

Our findings suggest that pedagogical friction is not a single intervention whose benefits increase simply as more friction is added. Rather, its value depends on how it is designed and where it occurs in the consent process. The designs with the strongest observed initial comprehension differed from the design with the strongest observed recovery after an initial attempt below the threshold, and these interventions carried different costs in time and perceived burden. At the same time, consent remained high even when participants did not demonstrate comprehension. We therefore interpret pedagogical friction as part of a broader consent process that can support initial understanding, identify misunderstanding, and create opportunities for re-review, while making the costs of those interventions visible. %can support demonstrated comprehension at different stages of the consent process, although the benefits are not uniform across conditions. The timed slide recap condition (G3) had the highest first-attempt threshold attainment, followed by the paced, sectioned highlighting condition (G4), while differences in mean quiz accuracy were more modest. Among participants who did not meet the threshold on the first attempt, G5 had the highest recovery measures among the retake conditions, although these comparisons are descriptive. These patterns suggest that supporting initial comprehension and supporting recovery after an initial attempt below the threshold are distinct design goals, with different forms of friction offering different benefits and costs.

\subsection{Pedagogical friction should be calibrated, not maximized}
Our first-attempt results do not support a simple relationship in which more friction consistently produces better comprehension. G3 had the highest observed first-attempt threshold attainment, followed by G4, while G5, which also included pacing and sectioning, had lower first-attempt performance. One difference between G3 and the other conditions is that G3 presented the key terms twice before the first quiz: once in the full policy text and again in simplified, illustrated form. G5 also included simplified language, but its blurbs explained individual highlighted passages in place, whereas the slides presented a condensed version of the key terms as a whole. One possible explanation for G3’s stronger observed first-attempt performance is that participants encountered the key terms a second time in a condensed format. Because repeated exposure, simplified visual presentation, and pacing were bundled in this condition, our data cannot determine which of these elements contributed to the observed pattern. The result nevertheless motivates future work testing whether a second, condensed encounter with consequential terms can support initial comprehension. 
%This suggests that a second, condensed encounter with the key terms may support initial comprehension more than added structure within a single pass. 

These comparisons are interpretive rather than causal. Mean accuracy differences across conditions were modest, and our conditions were designed incrementally, with several conditions adding forms of pedagogical friction relative to earlier conditions. We interpret the results primarily at the condition level rather than attributing observed differences to any single element.

For consent interface design, the important implication is not to maximize friction, but to select friction that serves a particular comprehension goal. Structured exposure and pacing are candidate design mechanisms for supporting attention to important terms, with our results reflecting the performance of the complete conditions in which these mechanisms appear. This distinction extends the broader friction-in-design view: the value of friction depends on what behavior the intervention is intended to support and where it appears in the interaction~\cite{frischmann2023friction,cox2016design}.

\subsection{Comprehension checks create opportunities for re-review}
The retake conditions provide a second design opportunity after a participant does not meet the comprehension threshold on their first attempt. 66.4\% of retakers improved their score on the second attempt. G5 showed the largest mean gain, the highest proportion of retakers who improved, and the highest second-attempt threshold attainment among the three retake conditions. Answer transitions also showed that retakers corrected previously missed questions more often than they introduced new errors.

These comparisons are descriptive because the retake analysis includes only participants who did not meet the threshold on their first attempt. We do not interpret the differences among G2, G3, and G5 retakers as isolated causal effects of individual interface components. Because retakers reviewed the policy material again and then answered the same questions, the gains cannot be attributed specifically to re-review rather than repeated testing or item familiarity. The retake conditions also differed in what re-review involved. In particular, G2 and G5 retakers returned to the policy text, while G3 retakers viewed only the slide recap and were not subject to the enforced per-slide minimum. G5's stronger observed recovery should therefore be interpreted in light of these differences in re-review design. The broader design implication is that a comprehension check can do more than classify whether a user passed or failed. It can identify when the comprehension threshold is not met and provide an opportunity for another review before the consent process continues. In this sense, demonstrated misunderstanding becomes a point of intervention within the consent process rather than simply an endpoint. This is consistent with prior work on embedded instruction, where support is delivered during the task at the point when the user can act on it~\cite{kumaraguru2007protecting}.

\subsection{Comprehension checks reveal a gap between agreement and demonstrated comprehension}
The consent decisions expose a limitation of treating agreement itself as evidence of understanding. In the ungated conditions, consent was nearly identical among participants who met the comprehension threshold (96.9\%) and those who did not (97.3\%).%96.9\% of participants who met the comprehension threshold and 97.3\% of those below the threshold chose to consent.

This result does not show that comprehension gating changes a person's willingness to consent. Participants below the threshold in the gated conditions were not allowed to provide consent, so their consent preferences under an ungated design are unknown. Instead, the finding shows that an ungated consent flow can record agreement even when the same interaction provides evidence that the user did not meet the comprehension threshold.

For interface design, comprehension checks can serve as a diagnostic component of consent rather than only as a gate. They can reveal when agreement and demonstrated comprehension are misaligned. Their value, therefore, is not necessarily that they prevent consent, but that they make visible a form of misalignment that a conventional consent interface leaves unobserved. This distinction is especially important for consequential consent interfaces, where a single affirmative action may otherwise be interpreted as stronger evidence of understanding than the interaction actually provides.

\subsection{The benefits of friction must be considered alongside its costs}
Pedagogical friction also introduces measurable interaction costs. The structured and paced conditions generally required more policy review time than the lighter conditions, and retakes added another round of review and assessment. Participants also reported greater burden in some conditions with more enforced exposure and retakes. At the same time, perceived burden did not increase uniformly across every form of friction, suggesting that the cost of an intervention depends on how the friction is implemented, not only on how much additional time it requires.

This creates a calibration problem for designers. A consent interface should provide enough support for comprehension without adding unnecessary burden to every user. Our results motivate an adaptive approach worth further investigation: provide structured support during the initial review, then reserve additional review and reassessment for users who do not initially demonstrate comprehension. %Our results suggest one possible approach: use structure and pacing during the initial review, then reserve additional review and reassessment for users who do not initially demonstrate comprehension. 
Such an approach could avoid imposing the full cost of repeated review on users who already meet the comprehension threshold while still providing a path for re-review for those who do not.

Our study does not determine an optimal amount of friction. The appropriate balance will depend on the importance of the terms, the consequences of misunderstanding them, and the burden imposed by the intervention. The broader contribution here is to treat these outcomes jointly: the value of pedagogical friction should be evaluated in terms of both the comprehension it supports and the costs it imposes. % with time and perceived burden instead of treating friction only as either a usability problem or a benefit.

\subsection{Selecting substantive terms for demonstrably informed consent}
\label{selecting_substantive_terms}
A practical challenge for demonstrably informed consent is deciding \emph{what} a user must understand~\cite{frischmann2026defendingconsent,frischmann2024better}. Privacy policies cover many topics, and what counts as a substantively important term can vary by context. In practice, the set of key terms is likely to reflect what a service expects users to notice, what matters for users making decisions, and what the service would later want to be able to enforce. Legal regulation may also determine what terms are substantively important. We therefore do not treat our quiz as a universal set of privacy terms that users must understand. Rather, it operationalizes a limited set of terms selected for this particular context. %is a concrete set chosen for this context to make teaching and assessing demonstrated comprehension feasible within a realistic consent flow.

In designing the study policy and quiz, we aimed for a stimulus that was concise enough to be read in an experiment while still reflecting common disclosure topics for an educational app. We centered the policy on GDPR and Children’s Online Privacy Protection Act (COPPA) relevant categories, including the types of personal information that may be collected, how that information may be used, and when it may be shared~\cite{gdprinfo_art13,ftc_coppa_faq,ecfr_coppa_rule_part312} (see Appendix Figure~\ref{fig:policy_fulltext}). The specific terms are less important than the broader design principle they illustrate: demonstrably informed consent requires an explicit decision about which terms are sufficiently consequential to warrant focused attention and assessment. %Our quiz items operationalize these categories as assessable terms: persistent identifiers (item~1), deletion rights and how parents can exercise them (item~2), geolocation collection and the conditions for verifiable parental consent (item~3), the COPPA age threshold for verifiable parental consent (item~5), and limits on sharing for business functions, legal compliance, and safety (item~6) (Appendix Table~\ref{tab:quiz_questions}). We also included a dispute resolution provision (item~4), because arbitration is a substantively important contractual term in many digital agreements even when it is not treated as a standard privacy disclosure~\cite{eisenberg2007arbitration,dasteel2017clickarbitration}.

Rather than treating a privacy policy as a single block of text that users are expected to understand uniformly, designers can identify a limited set of substantively important terms, % that are both decision-relevant and substantively important in the given context, then pair 
connect those terms to focused explanations or structured presentation, and assess comprehension. Pedagogical friction can then be targeted to the parts of the agreement where misunderstanding matters most, rather than applied uniformly across the entire policy. % with lightweight teaching mechanisms and aligned assessments. In our study, the interface interventions focus attention during policy review, while the quiz provides a direct signal of whether participants understood the selected terms. Future designs will need to choose which terms to emphasize, but the broader point remains that demonstrably informed consent requires not only disclosure, but also a deliberate selection of substantive terms and a way to verify demonstrated comprehension of them.

Taken together, our findings frame consent as an interaction process rather than a single recorded click. Such a process can identify substantively important terms, structure how those terms are presented, assess demonstrated comprehension, and provide opportunities for additional review when needed, while considering these interventions alongside their interaction costs. Pedagogical friction is therefore most useful not as a general strategy for slowing users down, but as a targeted design tool for supporting understanding at consequential points in the consent process.

\subsection{Limitations and future work}

Our study isolates consent design within a controlled experimental task. Participants were recruited to provide feedback on the app, so their consent decision was hypothetical rather than %Participants reviewed the privacy policy before directly browsing the app, and app exploration occurred only after the consent decision and was brief (at least two minutes). Participants also made the consent decision in a hypothetical scenario rather than 
related to their own child's actual use of the service. Likewise, their responses about their willingness to let their child use the app may reflect perceived usefulness or other considerations beyond privacy. %, and the study did not actually collect the personal data described in the policy. The 97.3\% figure characterizes the alignment between quiz performance and consent within our experimental task and should not be interpreted as an estimate of real world parental consent behavior. For the same reason, item~1 in Table~\ref{tab:exit_survey} reflects a broader post-review judgment about the app and may incorporate perceived usefulness or other non-privacy considerations.

Our study uses a short six-question quiz tied to a single policy and session, making the comprehension threshold necessarily coarse. %The six-question format makes the threshold necessarily coarse. 
We examined alternative thresholds of 4/6, 5/6, and 6/6 correct answers and found that observed patterns were similar at 4/6 and 5/6 but changed at 6/6. The 80\% threshold should be interpreted as an operational study criterion rather than a universal legal or design standard for informed consent. The quiz also measures immediate understanding of selected terms rather than deeper or lasting comprehension of policy implications. In addition, we did not directly measure cognitive load or mental effort with a validated scale, so the survey results should be interpreted as self-reported reactions rather than comprehensive measures of burden. Future work should use more fine-grained comprehension measures and examine how comprehension criteria should be calibrated across contexts and consequential terms.

Our experimental conditions bundle several design features, so the results should be interpreted primarily at the condition level rather than as estimates of the causal effects of individual interface components. %were designed incrementally, with several conditions adding forms of pedagogical friction relative to earlier conditions. Because some comparisons involve more than one design change, we interpret the results primarily at the condition level. 
In G3, for example, repeated exposure to key terms, simplified visual presentation, and enforced pacing cannot be disentangled. %, so its first-attempt advantage should not be attributed to any one of these. 
Comparisons among retakers are also descriptive because retakers are the subset of participants who did not meet the comprehension threshold on their first attempt and the retake conditions differed in what re-review entailed. In addition, our a priori power analysis targeted an omnibus comparison across six groups, so the study was not specifically powered for smaller pairwise or demographic subgroup differences. Future factorial studies could further examine the individual contributions of different forms of pedagogical friction, including the contribution of a second, condensed pass over key terms.

A further limitation concerns study completion. Our analyses include only participants who completed the study. If discontinuation was related to the amount or type of friction encountered, our estimates of perceived burden may underrepresent participants who were least tolerant of the interventions. Future studies should therefore examine abandonment and completion as outcomes when evaluating frictional consent interfaces.

Finally, our findings are based on one privacy policy, one children's educational application, and one population of U.S. parents. Replication across policies, domains, populations, and higher-stakes settings is needed to determine how broadly these patterns generalize.

\section{Conclusion}

We studied pedagogical friction as a design approach for supporting demonstrated comprehension in privacy consent. In a six-condition experiment with 293 parents, we found that different forms of pedagogical friction were associated with different stages of comprehension. Some designs showed stronger initial performance, while re-review provided an opportunity for many participants who initially fell below the comprehension threshold to improve. These benefits were accompanied by additional time and, in some conditions, greater perceived burden. At the same time, nearly all participants in ungated conditions chose to consent even when they did not demonstrate comprehension, underscoring the limits of treating recorded agreement as evidence of comprehension. Our findings suggest that consent is better understood as an interaction process in which designers can structure exposure, check comprehension, provide opportunities for re-review, and calibrate these interventions against their costs.

\section*{Ethics Statement}
Our study protocol was reviewed and approved by our IRB before recruitment. All participants were informed about the study procedure before they began and could stop at any time without penalty. Participants were paid \$18.75 per hour, prorated for time spent. We did not collect participant images, audio, or video. Participants were told that the app would access the types of data described in the privacy policy, but our study did not actually collect or access such personal data. In addition, to preserve study validity, we did not disclose in advance that the privacy policy consent itself was the focus of the experiment. At the end of the study, participants were debriefed about the study's focus and the limited disclosure used during the experiment.

Our study app stored study metadata on a secure server so that we could compute accuracy measures linked only to anonymous participant identifiers. Identifying information remained on the Positly platform and was not included in the analytic dataset or shared outside the research team.

%\bibliographystyle{ACM-Reference-Format}
\printbibliography

%\clearpage

\appendix

\section{Additional results}
\label{appendix_tables}

Table~\ref{tab:demographics} summarizes demographics for 293 participants. We report the distributions for gender, age, education, and household income.

\begin{table}[t]
\centering
\small
\caption{Participant demographics summary.}
\label{tab:demographics}
\begin{tabular}{
    @{}
    p{0.25\linewidth}
    >{\raggedleft\arraybackslash}p{0.04\linewidth}
    p{0.25\linewidth}
    >{\raggedleft\arraybackslash}p{0.04\linewidth}
    @{}
}
\toprule
\textbf{Gender} & \textbf{\%} &
\textbf{Age} & \textbf{\%} \\
\midrule
Female & 54.6 & 19--30 & 15.8 \\
Male   & 45.4 & 31--40 & 50.7 \\
       &      & 41--50 & 27.2 \\
       &      & 51--64 & 6.3 \\
\midrule
\textbf{Education} & \textbf{\%} &
\textbf{Household income} & \textbf{\%} \\
\midrule
High school or below & 30.9 &
Less than \$39{,}999 & 16.2 \\

Associate degree & 17.9 &
\$40{,}000 to \$59{,}999 & 15.9 \\

Bachelor's degree & 30.9 &
\$60{,}000 to \$84{,}999 & 21.7 \\

Graduate or professional degree & 20.3 &
\$85{,}000 to \$114{,}999 & 18.4 \\

& &
\$115{,}000 to \$149{,}999 & 15.9 \\

& &
\$150{,}000 or more & 11.9 \\
\bottomrule
\end{tabular}
\end{table}

Table~\ref{tab:demographics_outcomes} summarizes the study outcomes by age group and education level and supports the conclusion in \S~\ref{subsec:demo_outcomes}.

\begin{table}[t]
\centering
\small
\caption{Study outcomes by age group and education level.}
\label{tab:demographics_outcomes}
\begin{tabular}{
    >{\centering\arraybackslash}m{1cm}
    >{\centering\arraybackslash}m{1cm}
    >{\centering\arraybackslash}m{1cm}
    >{\centering\arraybackslash}m{1.5cm}
    >{\centering\arraybackslash}m{1.5cm}
    >{\centering\arraybackslash}m{1.5cm}
    >{\centering\arraybackslash}m{1.5cm}}
\toprule
\textbf{Factor} & \textbf{Category} & \textbf{Prop.} &
\textbf{\makecell{1st quiz\\mean acc.}} &
\textbf{\makecell{$\ge$80\% on\\1st quiz}} &
\textbf{\makecell{2nd quiz\\mean  acc.}} &
\textbf{\makecell{$\ge$80\% on\\2nd quiz}} \\
\midrule
\multirow{4}{*}{\textbf{Age}} 
& 19-30 & 15.8\% & 0.54 & 25.0\% & 0.51 & 26.7\% \\
& 31-40 & 50.7\% & 0.54 & 19.0\% & 0.59 & 26.2\% \\
& 41-50 & 27.2\% & 0.60 & 31.6\% & 0.66 & 25.0\% \\
& 51-64 &  6.3\% & 0.64 & 33.3\% & 0.52 & 0 \\
\midrule
\multirow{4}{*}{\textbf{Edu.}}
& \makecell{H.S.} & 30.9\% & 0.50 & 11.8\% & 0.57 & 26.3\% \\
& \makecell{Assoc.} & 17.9\% & 0.51 & 17.4\% & 0.57 & 22.2\% \\
& \makecell{BS} & 30.9\% & 0.52 & 21.1\% & 0.60 & 7.1\% \\
& \makecell{Grad.} & 20.3\% & 0.62 & 32.1\% & 0.50 & 0 \\
\bottomrule
\end{tabular}
\end{table}

Table~\ref{tab:time_cost_all} lists the time costs in median [Q1-Q3]. Q1 and Q3 are the 25th and 75th percentiles, providing additional context for the main findings in \S~\ref{subsec:time_cost}.

\begin{table}[t]
\centering
\caption{Time cost by group (seconds), reported as median [Q1--Q3], where Q1 and Q3 are the 25th and 75th percentiles. First-attempt times are computed over available timing observations; retake times are computed over retakers with available timing observations. In G3, policy time combines viewing of the plain-text policy and the slide recap, which were logged as a single duration. The seven slides carried an enforced 5-second minimum on the first attempt only, giving a 35-second floor on that portion.}
\label{tab:time_cost_all}
\small
\begin{adjustbox}{max width=\linewidth}
\begin{tabular}{
    >{\centering\arraybackslash}p{0.8cm} 
    >{\centering\arraybackslash}p{2cm}
    >{\centering\arraybackslash}p{2cm}
    >{\centering\arraybackslash}p{2.2cm} 
    >{\centering\arraybackslash}p{1.8cm}
    >{\centering\arraybackslash}p{1.8cm}
    >{\centering\arraybackslash}p{2.4cm}}
\toprule
\multirow{2}{*}{\textbf{Group}}  &
\multicolumn{3}{c}{\textbf{First attempt}} &
\multicolumn{3}{c}{\textbf{Second attempt}} \\
\cmidrule(lr){2-4}\cmidrule(lr){5-7}
& \textbf{Policy time} & \textbf{Quiz time} & \textbf{Total time}
& \textbf{Policy time} & \textbf{Quiz time} & \textbf{Total time} \\

\cmidrule(lr){1-7}
G0 & 10 [7-12]       & 115 [95.5-157]    & 122 [105-172]
   & --                    & --                      & -- \\
G1 & 8 [5.5-12]     & 112 [83.5-166.5]  & 121 [93-179]
   & --                    & --                      & -- \\
G2 & 39 [5-70]       & 65.5 [49-89]      & 117 [70-126]
   & 8.5 [5.5-68]     & 59 [44-104.75]     & 108.5 [46.5-155.75] \\
G3 & 70 [65.5-82.5]  & 80 [68.5-116]     & 152 [139-221.5]
   & 21 [15.5-48.5]   & 45 [33.5-72]       & 66 [53.5-135.5] \\
G4 & 55 [45-84]      & 88 [62-122]       & 162 [118-206]
   & --                    & --                      & -- \\
G5 & 51 [38.5-76.25] & 87 [67-104]       & 138 [115-184.25]
   & 80 [49.5-125]   & 65 [44-85.5]      & 143 [112.75-207.5] \\
\bottomrule
\end{tabular}
\end{adjustbox}
\end{table}

\section{Study Materials}

Table~\ref{tab:quiz_questions} lists the privacy policy quiz questions. The questions cover persistent identifiers (item~1), deletion rights and how parents can exercise them (item~2), geolocation collection and the conditions for verifiable parental consent (item~3), dispute resolution provision (item~4), the COPPA age threshold for verifiable parental consent (item~5), and limits on sharing for business functions, legal compliance, and safety (item~6).

\begin{table}[t]
\centering
\caption{Privacy policy quiz questions.}
\label{tab:quiz_questions}
\renewcommand{\arraystretch}{1.15}
\small
\begin{adjustbox}{max width=\linewidth}

\begin{tabular}{
    >{\centering\arraybackslash}p{0.4cm}
    >{\raggedright\arraybackslash}p{4.5cm}
    >{\raggedright\arraybackslash}p{7cm}
}
\toprule
\textbf{Item} & \textbf{Question} & \textbf{Answer choices} \\
\midrule
1 &
Which of the following is an example of a ``persistent identifier'' KidsZone may collect from a child's device? &
\begin{tabular}[t]{@{}l@{}}
(A) Social Security Number \\
(B) Medical history \\
(C) Web browser type or IP address \\
(D) Physical home address \\
(E) All of the above \\
(F) None of the above
\end{tabular}
\\
\midrule
2 &
How can parents delete the personal information KidsZone has collected from their child's account? &
\begin{tabular}[t]{@{}l@{}}
(A) By reporting the child's profile to moderators \\
(B) By logging in to the child's account or contacting\\KidsZone to make the request \\
(C) By clearing their internet browser history \\
(D) By disabling the app from their phone
\end{tabular}
\\
\midrule
3 &
What does KidsZone do before collecting geolocation data that could reveal a child's street address? &
\begin{tabular}[t]{@{}l@{}}
(A) Automatically masks it before storing \\
(B) Requests approval from the app store \\
(C) Seeks verifiable parental consent before collection \\
(D) Only collects it from children over the age of 10
\end{tabular}
\\
\midrule
4 &
If a dispute arises related to KidsZone's privacy policy, how is it resolved? &
\begin{tabular}[t]{@{}l@{}}
(A) Through a public court hearing in the child's home state \\
(B) A binding decision is made outside of court by a third party \\
(C) By a vote from the KidsZone moderation team \\
(D) By suspending the child's account until further notice
\end{tabular}
\\
\midrule
5 &
Which of the following best describes when verifiable parental consent is required under COPPA for collecting personal information online? &
\begin{tabular}[t]{@{}l@{}}
(A) For any individual under the age of 18 using a U.S.-based\\website \\
(B) Only if the website knows the user is under 10 years old \\
(C) For users aged 13-17 if they are not in school \\
(D) For children under 13 years old before collecting personal\\information
\end{tabular}
\\
\midrule
6 &
Under what conditions may KidsZone share a child's personal information with others? &
\begin{tabular}[t]{@{}l@{}}
(A) When a parent publicly shares the child's account details \\
(B) Only if the child wins a prize \\
(C) To perform business functions, comply with laws, or\\protect safety \\
(D) When a third-party company requests it for advertising
\end{tabular}
\\
\bottomrule
\end{tabular}
\end{adjustbox}
\end{table}

Figure~\ref{fig:policy_fulltext} lists the full privacy policy terms used for our study. We centered the terms on GDPR- and COPPA-relevant categories, including the types of personal information that may be collected, how that information may be used, and when it may be shared.

\begin{figure}[p]
\centering

\setlength{\fboxsep}{3pt}
\setlength{\fboxrule}{0.5pt}

\fbox{%
\begin{minipage}{0.96\textwidth}

\scriptsize
\setlength{\parskip}{0pt}
\setlength{\baselineskip}{7pt}

{\centering \textbf{\normalsize KidsZone Privacy Policy}\par}
\vspace{2pt}

KidsZone is committed to protecting the privacy of children who use our site. The U.S. Children's Online Privacy Protection Act ("COPPA") requires parental consent for collecting personal information from children, for all children under the age of 13 years old.\par

{\centering\textbf{1. THE INFORMATION WE COLLECT FROM CHILDREN, HOW WE USE IT, AND HOW AND WHEN WE COMMUNICATE WITH PARENTS}\par}

KidsZone offers to its users a range of sites and applications, some targeted at children. Activities may collect information from children. We retain child-collected personal data only as long as necessary for the activity, security, or legal requirements. If information is collected inconsistent with COPPA, we will delete it or seek parental consent.\par

\textbf{1.1 Registration}

Children can register to view content, play games, and participate in contests. During registration, we may collect parent email, child’s first name, gender, username, password, and birth date to validate age. We don’t ask for more information than necessary for the activity.

\textbf{1.2 About the collection of parent email address}

Before collecting any personal data from children under 13 years old, we ask for a parent/guardian email address so we can obtain parental consent. If you believe your child’s data was collected without notice, contact privacycontact@kidszone.com. Parent emails for consent won’t be used for marketing unless opted in.

\textbf{1.3 Verifiable Parental Consent}

In the event KidsZone wishes to collect personal information from a child, COPPA requires that we first seek a parent or guardian’s consent by email. In the email we will explain what information we are collecting, how we plan to use it, how the parent can provide consent, and how the parent can revoke consent. If we do not receive parental consent within a reasonable time, we will delete the parent contact information and any other information collected from the child in connection with that activity.

\textbf{1.4 Content Generated by a Child}

Certain activities on our sites and applications allow children to create or manipulate content and save it with KidsZone. Some of these activities do not require children to provide any personal information and therefore do not require parental consent. If an activity potentially allows a child to insert personal information in their created content, we will either pre-screen the submission to delete any personal information, or we will seek verifiable parental consent by email for the collection. Examples of created content that may include personal information are stories or other open-text fields, and drawings that allow text or free-hand entry of information. If, in addition to collecting content that includes personal information, KidsZone also plans to post the content publicly or share it with a third party for the third party’s own use, we will seek verifiable parental consent by email.

\textbf{1.5 Email Contact with a Child}

In connection with certain activities or services, we may collect a child’s online contact information, such as an email address, in order to communicate with the child more than once. In such instances we will retain the child’s online contact information to honor the request and for no other purpose such as marketing. One example would be a newsletter that provides occasional updates about a site, game/activity, television show, personality/character, or feature movie. Whenever we collect a child’s online contact information for ongoing communications, we will require a parent email address in order to notify the parent about the collection and use of the child’s information, as well as to provide the parent an opportunity to prevent further contact with the child. On some occasions a child may be engaged in more than one ongoing communication, and a parent may be required to “opt-out” of each communication individually.

\textbf{1.6 Geolocation Data}

If a child-directed KidsZone site or application collects geolocation information that is specific enough to equate to the collection of a street address, we will first seek parental consent via email.

\textbf{1.7 Persistent Identifiers}

When children interact with us, certain information may automatically be collected, both to make our sites and applications more interesting and useful to children and for various purposes related to our business. Examples include the type of computer operating system, the child’s IP address or mobile device identifier, the web browser, the frequency with which the child visits various parts of our sites or applications, and information regarding the online or mobile service provider. This information is collected using technologies such as cookies, flash cookies, web beacons, and other unique identifiers. This information may be collected by KidsZone or by a third party. This data is principally used for internal purposes only, in order to:

- provide children with access to features and activities on our sites and applications

- conduct research and analysis to address the performance of our sites and applications

- generate anonymous reporting for use by KidsZone

KidsZone may share or disclose a child's personal information as needed to provide our Service or with your consent or permission. For example, we share information with our trusted vendors, third party service providers and individuals to provide services for us on our behalf, which may include analytics providers and hosting services. We may also share personal information if we have a good faith belief that access, use, preservation, or disclosure of such information is reasonably necessary to (a) satisfy any applicable law, regulation, legal process, or enforceable governmental request; (b) enforce applicable Terms of Service, including investigation of potential violations thereof; (c) detect, prevent or otherwise address fraud, security or technical issues; (d) protect the rights, property, or personal safety of KidsZone, its users, or the public; or (e) as required or permitted by law If KidsZone becomes involved in a merger, acquisition, bankruptcy, change of control, or any form of sale of some or all of its assets, personal information may be transferred or disclosed in connection with the business transaction, subject to any applicable laws. We may also share aggregate or de-identified information in a manner that cannot be reasonably used to identify an individual user.

{\centering\textbf{2. PARENTAL CHOICES AND CONTROLS}\par}

At any time, parents can refuse to permit us to collect further personal information from their children in association with a particular account, and can request that we delete from our records the personal information we have collected in connection with that account. Please keep in mind that a request to delete records may lead to a termination of an account, membership, or other service. Where a child has registered for a KidsZone account, we use the following method to allow parents to access, change, or delete the personally identifiable information that we have collected from their children. Parents can ask for deletion of all personal information collected in connection with their child's account and KidsZone will delete this information. Any other inquiries may be directed to: 

KidsZone Phone: (000) 123-4567 Email: privacycontact@kidszone.com

In any correspondence such as e-mail or mail, please include the child’s username and the parent’s email address and telephone number. To protect children’s privacy and security, we will take reasonable steps to help verify a parent’s identity before granting access to any personal information.

{\centering\textbf{3. ARBITRATION of DISPUTES}\par}

In the event a dispute shall arise between the parties to this policy, it is hereby agreed that the dispute shall be referred to United States Arbitration and Mediation for arbitration in accordance with United States Arbitration and Mediation Rules of Arbitration. “Dispute” includes any claim, dispute, action, or other controversy, whether based on past, present, or future events, whether based in contract, tort, statute, or common law, between you and KidsZone. The arbitrator’s decision shall be final and binding.

\end{minipage}}
\caption{Privacy policy used as the study stimulus.}
\label{fig:policy_fulltext}
\Description{Full text of the KidsZone privacy policy used as the study stimulus.}
\end{figure}

\section{Experimental Interfaces}

\begin{figure}[t]
  \centering  \includegraphics[width=\linewidth]{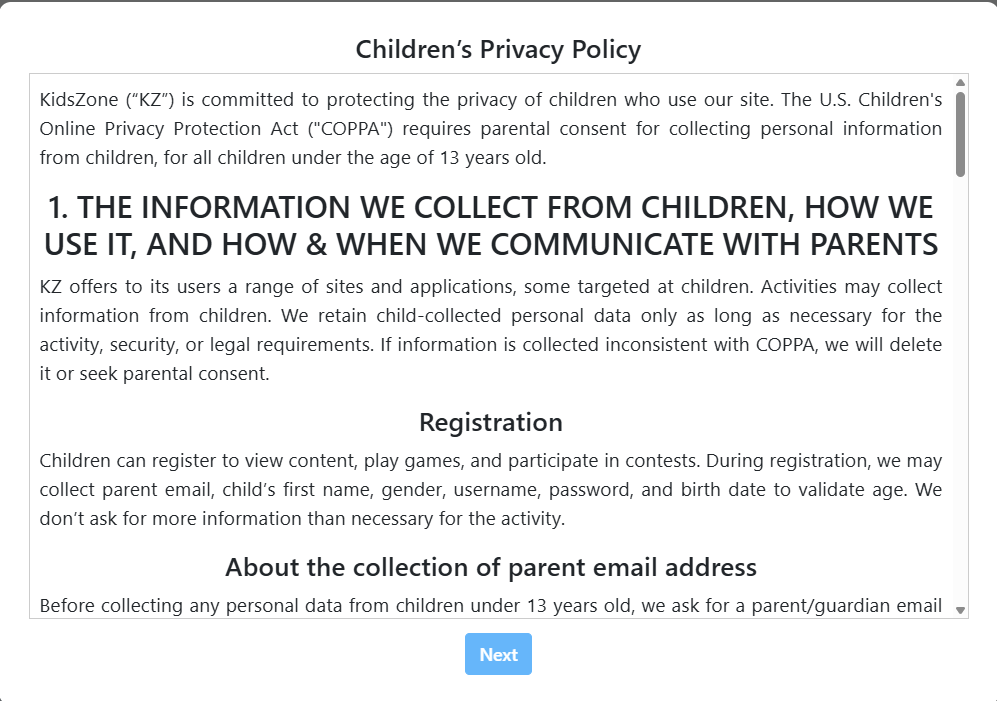}
  \caption{G0 plain text privacy policy review interface.}
  \label{fig:g0_policy}
  \Description{G0 plain text privacy policy review interface.}
\end{figure}

\begin{figure}[t]
  \centering  \includegraphics[width=\linewidth]{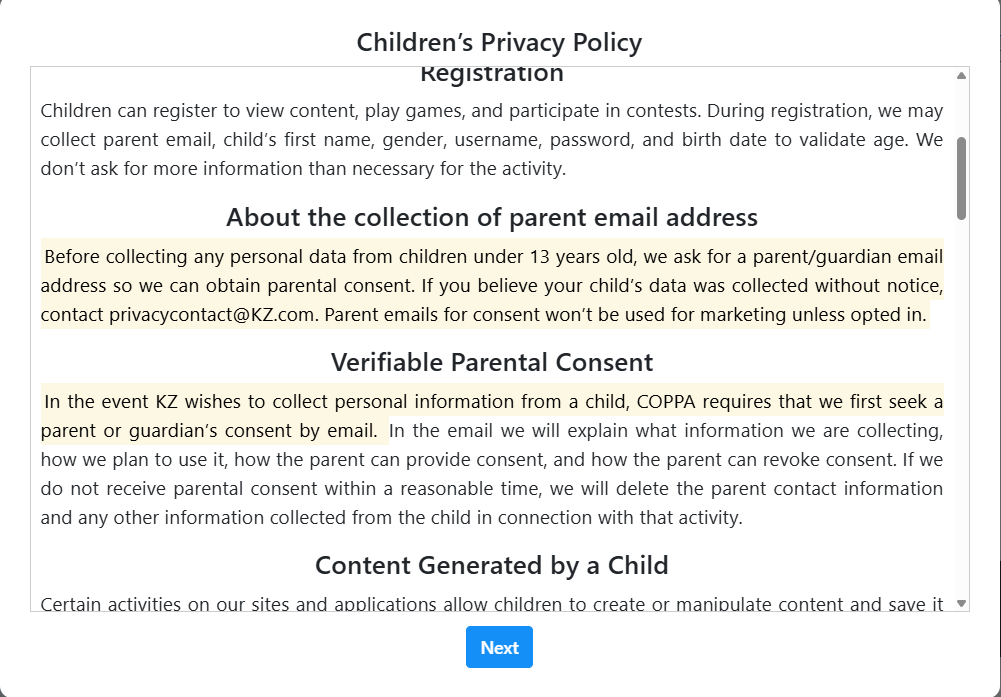}
  \caption{G1 highlighted privacy policy review interface.}
  \label{fig:g1_policy}
  \Description{G1 highlighted privacy policy review interface.}
\end{figure}

\begin{figure}[t]
  \centering  \includegraphics[width=\linewidth]{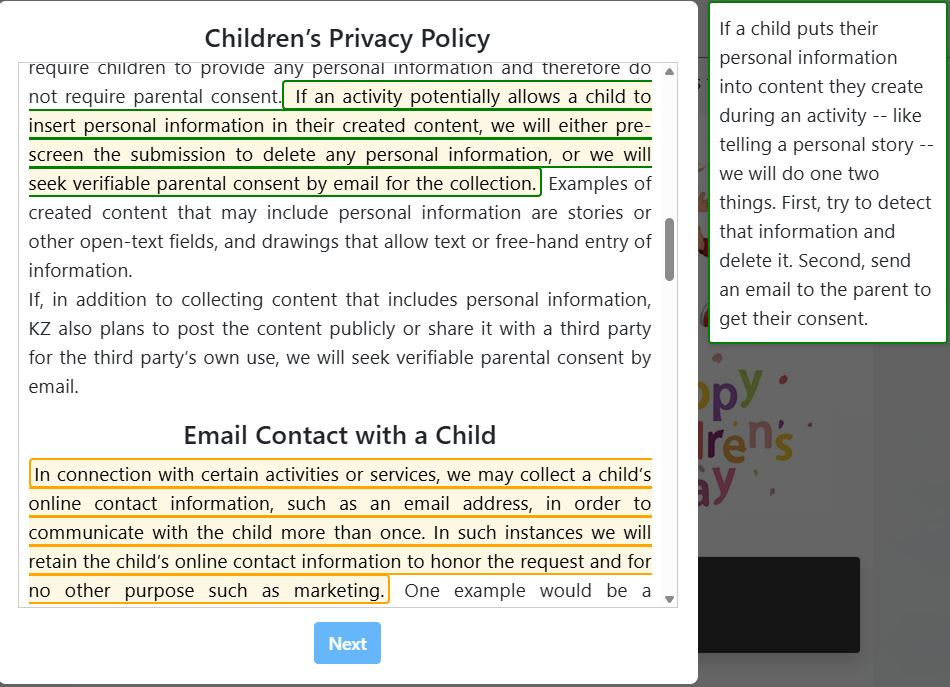}
  \caption{G2 highlighted privacy policy review interface with explanatory blurbs.}
  \label{fig:g2_policy}
  \Description{G2 highlighted privacy policy review interface with explanatory blurbs.}
\end{figure}

\begin{figure}[t]
  \centering

  \begin{minipage}{\linewidth}
    \centering
    \includegraphics[width=\linewidth]{G0-policy.png}
  \end{minipage}

  \begin{minipage}{\linewidth}
    \centering
    \includegraphics[width=0.82\linewidth]{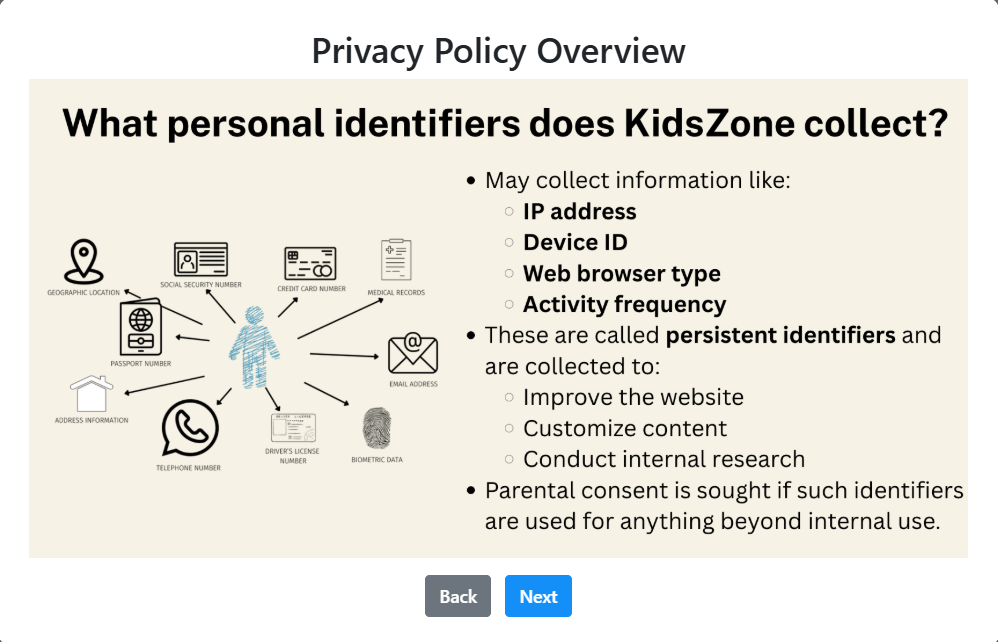}
  \end{minipage}

  \caption{G3 timed slide recap policy review interface: full policy page (top) and a sample timed slide recap (bottom).}
  \label{fig:g3_policy}

  \Description{G3 timed slide recap privacy policy review interface.}
\end{figure}

\begin{figure}[t]
  \centering  \includegraphics[width=\linewidth]{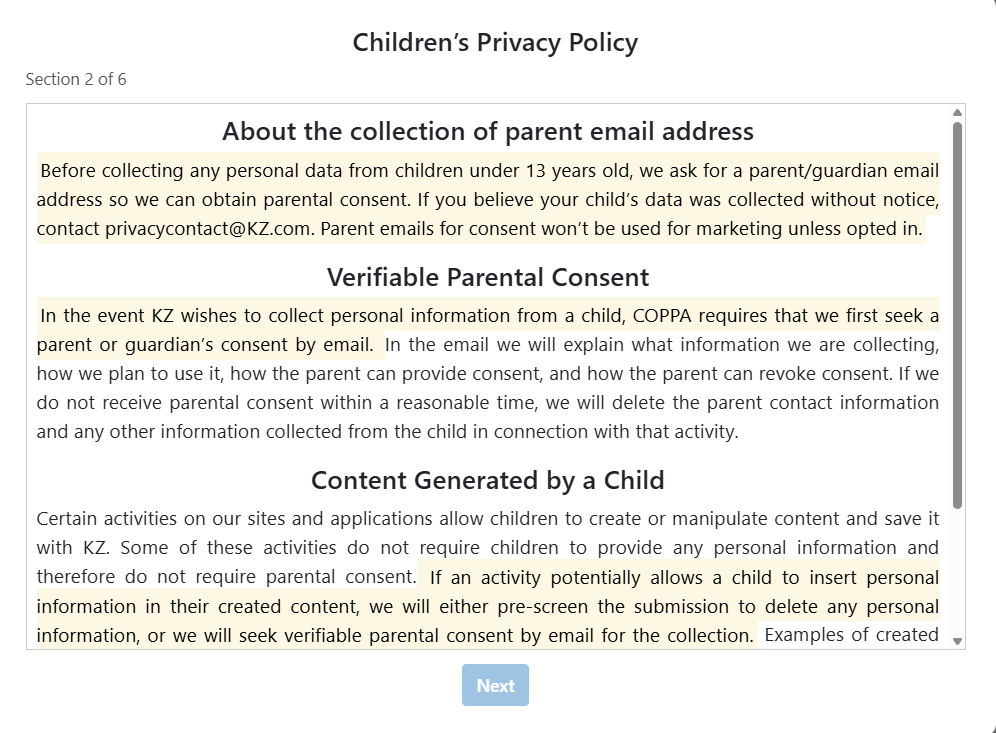}
  \caption{G4 sectioned highlighted privacy policy review interface with pacing.}
  \label{fig:g4_policy}
  \Description{G4 sectioned highlighted privacy policy review interface with pacing.}
\end{figure}

\begin{figure}[t]
  \centering  \includegraphics[width=\linewidth]{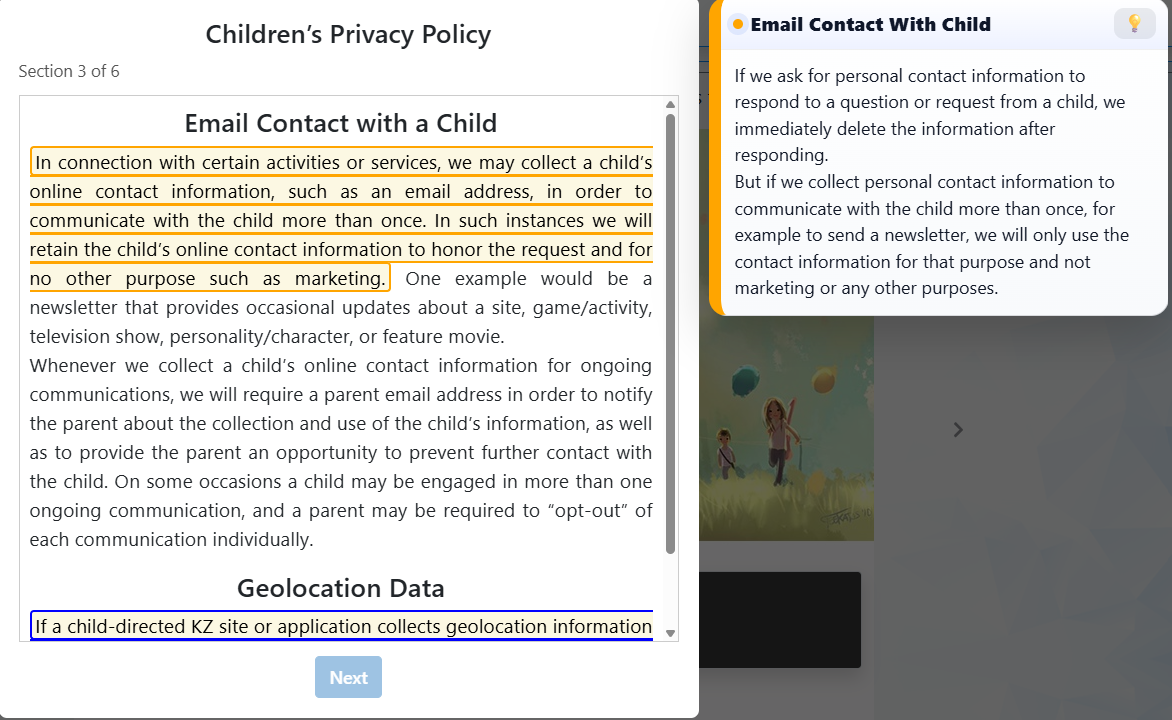}
  \caption{G5 sectioned highlighted privacy policy review interface with blurbs and pacing.}
  \label{fig:g5_policy}
  \Description{G5 sectioned highlighted privacy policy review interface with blurbs and pacing.}
\end{figure}

Figures~\ref{fig:g0_policy}-\ref{fig:g5_policy} show full-size examples of the privacy policy review interfaces for the six conditions, with the specific condition friction elements visible.

\section{Example Learning Pages}
\label{}

\begin{figure}[t]
  \centering  \includegraphics[width=\linewidth]{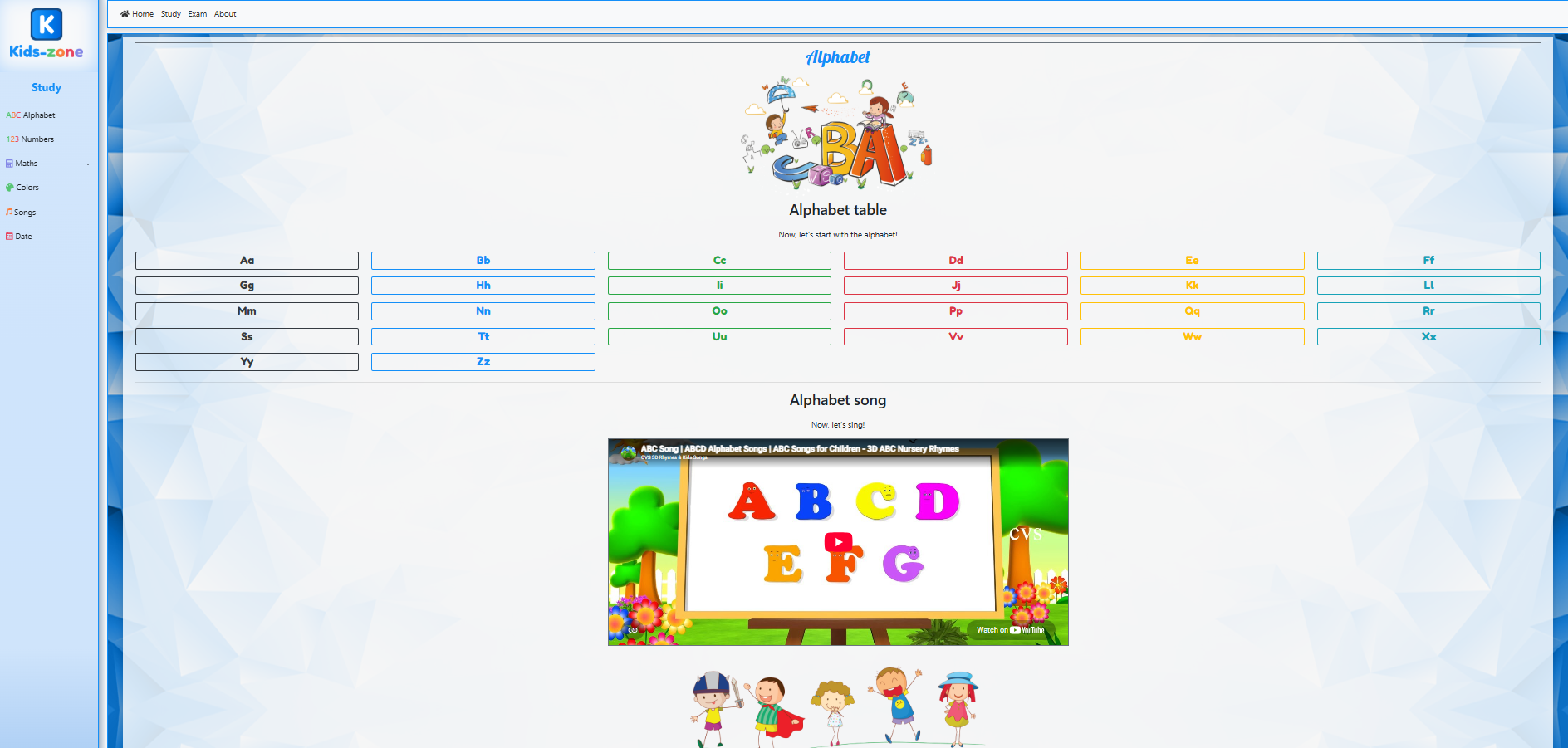}
  \caption{KidsZone alphabet learning page.}
  \label{fig:kz_alphabet}
  \Description{KidsZone alphabet learning page.}
\end{figure}

\begin{figure}[t]
  \centering  \includegraphics[width=\linewidth]{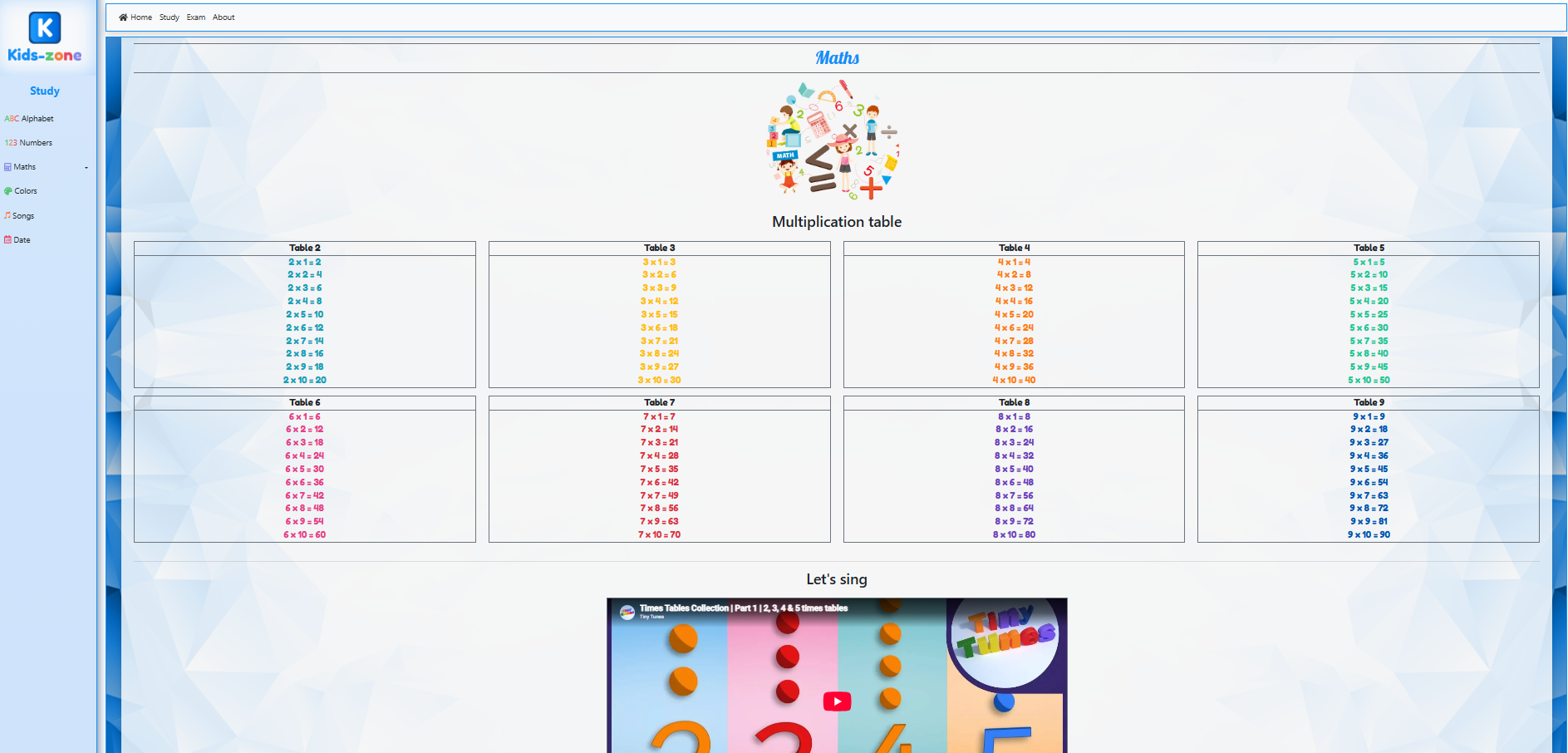}
  \caption{KidsZone multiplication practice page.}
  \label{fig:kz_math}
  \Description{KidsZone multiplication practice page.}
\end{figure}

\begin{figure}[t]
  \centering  \includegraphics[width=\linewidth]{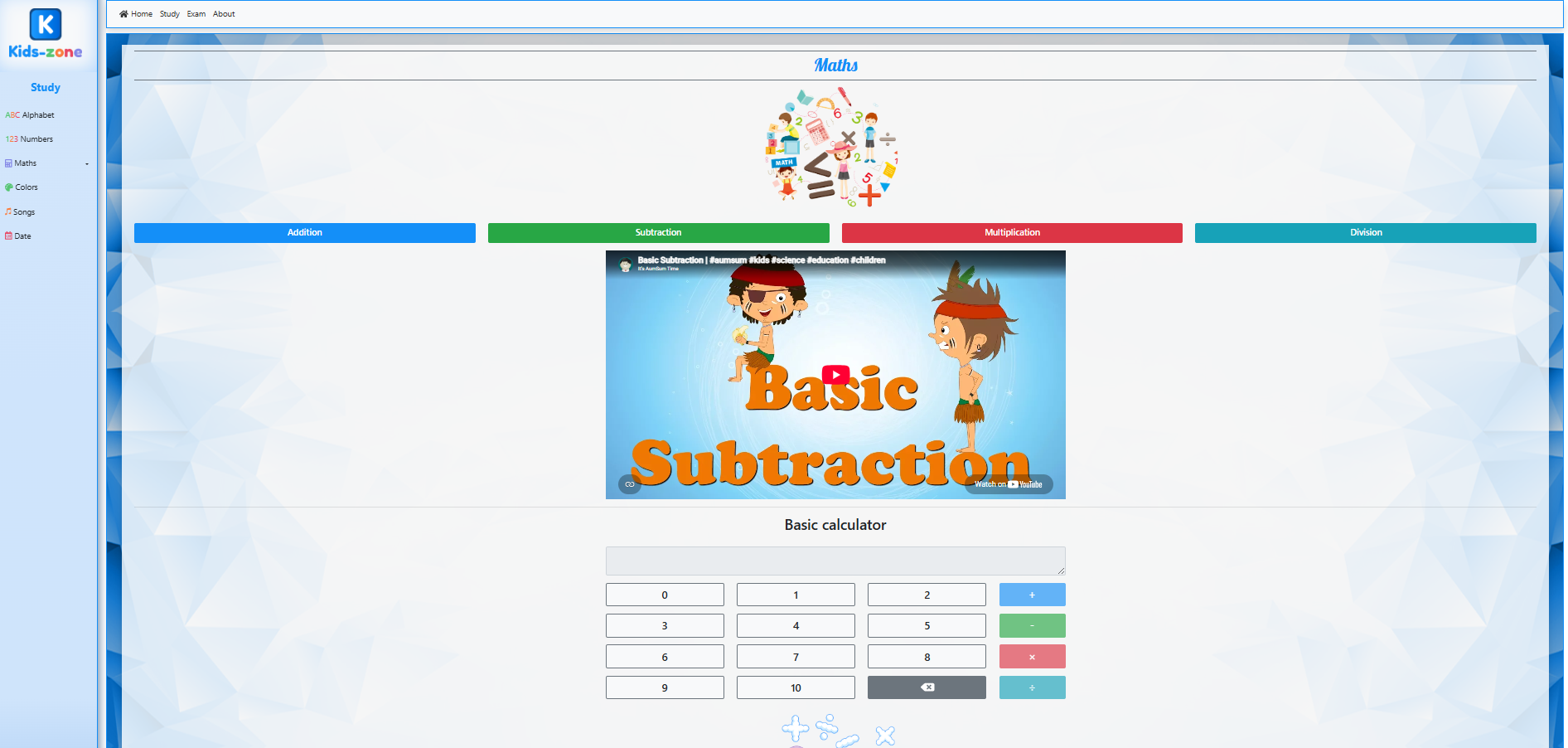}
  \caption{KidsZone subtraction practice page.}
  \label{fig:kz_sub}
  \Description{KidsZone subtraction practice page.}
\end{figure}

Figures~\ref{fig:kz_alphabet}-\ref{fig:kz_sub} show pages that teach children basic skills, such as the alphabet and simple math practice.

\end{document}